\documentclass[aps,amsmath,amssymb,floatfix,twocolumn]{revtex4}

\usepackage{bm}
\usepackage{graphicx}
\usepackage{amsbsy}
\usepackage{dcolumn}
\usepackage{color}

\newcommand{\be}{\begin{equation}}
\newcommand{\ee}{\end{equation}}
\newcommand{\bea}{\begin{eqnarray}}
\newcommand{\eea}{\end{eqnarray}}

\begin{document}

\title{Tensor Fermi liquid parameters in nuclear matter \\ from chiral effective field theory}

\author{J.\ W.\ Holt$^1$, N.\ Kaiser$^2$ and T.\ R.\ Whitehead$^1$}
\affiliation{$^1$Cyclotron Institute and Department of Physics and Astronomy, Texas 
A\&M University, College Station, TX}
\affiliation{$^2$Physik Department, Technische Universit\"{a}t M\"{u}nchen,
    D-85747 Garching, Germany}

\begin{abstract}

We compute from chiral two- and three-body forces the complete quasiparticle interaction
in symmetric nuclear matter up to twice nuclear matter saturation density. 
Second-order perturbative contributions that account for Pauli-blocking and medium
polarization are included, allowing for an exploration of the full set of central and
noncentral operator structures permitted by symmetries and the long-wavelength
limit. At the Hartree-Fock level, the next-to-next-to-leading order three-nucleon force 
contributes to all noncentral interactions, and their strengths grow approximately linearly
with the nucleon density up that of saturated nuclear matter. Three-body forces 
are shown to enhance the already strong 
proton-neutron effective tensor interaction, while the corresponding like-particle tensor
force remains small. We also find a large isovector cross-vector interaction but small
center-of-mass tensor interactions in the isoscalar and isovector channels. 
The convergence of the expansion of the noncentral quasiparticle interaction in Landau 
parameters and Legendre polynomials is studied in detail.

\end{abstract}

\maketitle


\section{Introduction}

Fermi liquid theory \cite{landau57a,landau57b,landau59,migdal64} is widely used to 
describe the transport, response and dynamical properties of nuclear many-body systems \cite{brown71,iwamoto82,haensel82,backman85,wambach93,friman96,benhar07,pastore12,benhar13}. 
The key quantity in this theory is the quasiparticle interaction, defined as the second functional
derivative of the energy with respect to the quasiparticle distribution function. 
For many years the primary focus of investigation has been the central part of the quasiparticle 
interaction and its associated Fermi liquid parameters, which are directly related to 
static properties of the interacting ground state such as the incompressibility, isospin-asymmetry
energy and magnetic susceptibility. The central terms include scalar operators in spin and 
isospin space, but more recently noncentral contributions \cite{haensel75,schwenk04} that 
couple spin and momenta have been studied together with 
their impact on the density and spin-density response functions of neutron matter
\cite{olsson04,bacca09,pethick09,davesne15prc}. Extending these results to nuclear matter
with equal numbers of protons and neutrons and to systems with arbitrary isospin 
asymmetry will be needed to better understand neutrino transport and 
emissivity in neutron stars, proto-neutron star cooling \cite{roberts12prl}, electron transport 
in neutron stars \cite{bertoni15}, the evolution of shell structure and single-particle
states in nuclei far from stability \cite{otsuka05,otsuka06}, and nuclear 
collective excitations (spin and spin-isospin 
modes together with rotational modes of deformed nuclei) \cite{friman81,cao09,co12}.

An important motivation of the present work is to provide microscopic guidance for the 
tensor forces employed in modern mean field effective interactions and nuclear energy 
density functionals. Including as well new estimates and uncertainties on the central
Fermi liquid parameters, which are more directly related to nuclear observables, the 
present study will complement other recent efforts 
\cite{pudliner96,brown14,roggero15,buraczynski16,rrapaj16,zhang18} to constrain
 energy density functionals 
from microscopic many-body theory. The importance of tensor forces
in mean field modeling is a question of ongoing debate.
While there is skepticism \cite{lesinski07,lalazissis09} 
that tensor forces can lead to a meaningful 
improvement in fits to nuclear ground state energies, there is strong evidence that 
the description of single-particle energies \cite{otsuka05,otsuka06,colo07,zuo08}, 
beta-decay half-lives \cite{minato13}, and spin-dependent collective excitations 
\cite{cao09,co12} are systematically improved with the inclusion of tensor forces 
(for a recent review, see Ref.\ \cite{sagawa14}). One of the main driving questions is 
the extent to which the effective medium-dependent tensor force in mean field models 
resembles the fundamental tensor component of the free-space nucleon-nucleon 
interaction arising from $\pi$ + $\rho$ meson exchange. A main conclusion of the 
present work is that the proton-neutron effective tensor force is enhanced over the 
free-space tensor interaction due to three-body forces and second-order 
perturbative contributions. On the other hand, the proton-proton and neutron-neutron
tensor forces are considerably smaller in magnitude. In addition, we find evidence for a 
large isovector cross-vector interaction that to our knowledge has not been previously
studied in phenomenological mean field modeling of nuclei. 

The quasiparticle interaction can be computed microscopically from realistic two- and
three-body forces starting from the perturbative expansion of the energy density
and taking appropriate functional derivatives with respect to the Fermi distribution functions.
For nuclear or astrophysical systems with densities near or above that of saturated nuclear matter, it is
essential to consider the effects of three-body forces. To date three-nucleon forces have been 
included in calculations of the central and exchange-tensor quasiparticle interaction in 
nuclear matter \cite{brown77,shen03,zuo03,kaiser06,gambacurta11,holt12npa} and the full quasiparticle 
interaction in neutron matter \cite{holt13prca}. In the present work our aim is to extend the 
calculations in Ref.\ \cite{holt13prca} to the case of symmetric nuclear matter. This is a natural 
step before considering the more general case of isospin-asymmetric nuclear
matter.

We take as a starting point a class 
\cite{entem03,coraggio07,marji13,coraggio13,coraggio14,sammarruca15} 
of realistic two and three-body nuclear forces derived within the 
framework of chiral effective field theory \cite{weinberg79,epelbaum09rmp,machleidt11}. 
The two-body force is treated at both next-to-next-to-leading order (N2LO) and N3LO
in the chiral power counting, while the three-body force is only considered at N2LO.
Although the inclusion of consistent three-body forces at N3LO \cite{bernard08,bernard11} 
in the chiral power counting will be needed for improved theoretical uncertainty estimates 
\cite{tews13,drischler16,drischler17}, the present set of nuclear force models has been
shown to give a good 
description of nuclear matter saturation \cite{coraggio14,holt17prc}, the nuclear liquid-gas 
phase transition \cite{wellenhofer14}, and the volume components of nucleon-nucleus optical 
potentials \cite{holt13prcb,holt16prc} when used at second order in many-body perturbation theory.
In addition to the order in the chiral expansion, the resolution scale (related to the 
momentum-space cutoff in the nuclear potential) is varied in order to assess the theoretical 
uncertainties. 

The paper is organized as follows. In Section \ref{qpsnm} we review the derivation of the
quasiparticle interaction and associated Fermi liquid parameters from microscopic nuclear two- and 
three-body interactions. We present a general method to extract the central and noncentral components
of the quasiparticle interaction from appropriate linear combinations of spin- and isospin-space
matrix elements. We also benchmark the numerical calculations of the second-order 
contributions to the quasiparticle interaction against semi-analytical results for model interactions 
of one-boson exchange type. In Section \ref{results} we present analytical expressions 
for the Landau parameters arising from the 
leading N$^2$LO chiral three-body force together with numerical results for the second-order
contributions from two- and three-body forces.
We end with a summary and conclusions.


\section{Quasiparticle interaction in symmetric nuclear matter}
\label{qpsnm}
\subsection{General structure of the quasiparticle interaction}
\label{gs}

The quasiparticle interaction in symmetric nuclear matter has the general form
\cite{schwenk04}
\begin{equation}
{\cal F}(\vec p_1, \vec p_2\,) = {\cal A }(\vec p_1, \vec p_2\,) + 
{\cal A^\prime }(\vec p_1, \vec p_2\,) \vec \tau_1 \cdot \vec \tau_2,
\end{equation}
where 
\begin{eqnarray}
{\cal A}(\vec p_1, \vec p_2\,) &=& f(\vec p_1, \vec p_2\,) + g(\vec p_1, 
\vec p_2\,) \vec \sigma_1 \cdot \vec \sigma_2 \nonumber \\
&&\hspace{-.5in}+ h (\vec p_1, \vec p_2\,) 
S_{12}(\hat q) + k (\vec p_1, \vec p_2\,) S_{12}(\hat P) \nonumber \\ 
&& \hspace{-.5in}+\ell (\vec p_1, \vec p_2\,) (\vec \sigma_1 \times \vec \sigma_2)\cdot 
(\hat q \times \hat P),
\label{qpi}
\end{eqnarray}
and analogously for ${\cal A^\prime}$ except with the replacement 
$\{f,g,h,k,\ell\} \longrightarrow \{f^\prime,g^\prime,h^\prime,k^\prime,\ell^\prime \}$.
The relative momentum is defined by $\vec q = \vec p_1 - \vec p_2$ and
the center of mass momentum is given by $\vec P = \vec p_1 + \vec p_2$. 
The tensor operator has the usual form
$S_{12}(\hat v) = 3 \vec \sigma_1 \cdot \hat v\, \vec \sigma_2 \cdot \hat 
v -\vec \sigma_1 \cdot\vec \sigma_2$. 
The interaction in Eq.\ (\ref{qpi}) is invariant under rotations, time-reversal, parity, and 
the exchange of particle labels. The presence of the medium breaks Galilean 
invariance, and two new structures (the ``center-of-mass tensor'' 
$S_{12}(\hat P)$ and ``cross-vector'' $(\vec \sigma_1 \times \vec \sigma_2)\cdot 
(\hat q \times \hat P)$ operators) arise \cite{schwenk04} that depend explicitly on the 
center-of-mass momentum $\vec P$. Neither of these terms are found in the 
free-space nucleon-nucleon potential.

By assumption the two quasiparticle momenta 
$\vec p_1$ and $\vec p_2$ lie on the Fermi surface ($|\vec p_1| = |\vec p_2| =k_f$) 
and therefore the scalar 
functions $\{f,g,h,k,\ell,f^\prime,g^\prime,h^\prime,k^\prime,\ell^\prime \}$
admit a decomposition in Legendre polynomials:
\begin{eqnarray}
f({\vec p}_1,{\vec p}_2) &=& \sum_{L=0}^\infty f_L(k_f) P_L(\mbox{cos } \theta),\nonumber \\
f^\prime({\vec p}_1,{\vec p}_2) &=& \sum_{L=0}^\infty f^\prime_L(k_f) P_L(\mbox{cos } \theta), \dots
\label{gflp}
\end{eqnarray}
where $\cos \theta ={\hat p}_1 \cdot {\hat p}_2$ and $q = 2k_f\, {\rm sin}\, (\theta /2)$
and $P=2k_f \cos(\theta/2)$.
The expansion coefficients $f_L, f^\prime_L,\dots$ are referred to as the Fermi liquid
parameters. In relating the Fermi liquid parameters to nuclear observables, it is often
convenient to multiply by the density of states 
\begin{equation}
N_0 = 2 M^* k_f / \pi^2
\label{dos}
\end{equation}
with $M^*$ the effective nucleon mass, to obtain 
dimensionless parameters $F_L = N_0 f_L,\dots$.

The ten scalar functions in Eq.\ (\ref{qpi})
can be extracted from linear combinations of the spin-space and isospin-space matrix elements, but
the decomposition will depend on the orientation of the orthogonal vectors $\vec q$ and
$\vec P$. For instance, if $\vec P = P \hat z$ and $\vec q = q \hat x$, then
\begin{eqnarray}
f                   &&=(6{\cal F}_{1,1;1,1}^1+3{\cal F}_{1,0;1,0}^1+3{\cal F}_{0,0;0,0}^1 \nonumber \\
&&\hspace{.1in}+2{\cal F}_{1,1;1,1}^0+{\cal F}_{1,0;1,0}^0+{\cal F}_{0,0;0,0}^0)/16\,, \nonumber \\
f^\prime &&=(2{\cal F}_{1,1;1,1}^1+{\cal F}_{1,0;1,0}^1+{\cal F}_{0,0;0,0}^1 \nonumber \\
&&\hspace{.09in}-2{\cal F}_{1,1;1,1}^0-{\cal F}_{1,0;1,0}^0-{\cal F}_{0,0;0,0}^0)/16\,, \nonumber \\
g            &&=(6{\cal F}_{1,1;1,1}^1+3{\cal F}_{1,0;1,0}^1-9{\cal F}_{0,0;0,0}^1 \nonumber \\
&&\hspace{.1in}+2{\cal F}_{1,1;1,1}^0+{\cal F}_{1,0;1,0}^0-3{\cal F}_{0,0;0,0}^0)/48\,, \nonumber \\
g^\prime&&=(2{\cal F}_{1,1;1,1}^1+{\cal F}_{1,0;1,0}^1-3{\cal F}_{0,0;0,0}^1 \nonumber \\
&&\hspace{.09in}-2{\cal F}_{1,1;1,1}^0-{\cal F}_{1,0;1,0}^0+3{\cal F}_{0,0;0,0}^0)/48\,, \nonumber \\
h            &&=(3{\cal F}_{1,1;1,-1}^1+{\cal F}_{1,1;1,-1}^0)/12\,, \nonumber \\
h^\prime&&=({\cal F}_{1,1;1,-1}^1-{\cal F}_{1,1;1,-1}^0)/12\,, \nonumber \\
k            &&=(3{\cal F}_{1,1;1,1}^1-3{\cal F}_{1,0;1,0}^1+3{\cal F}_{1,1;1,-1}^1 \nonumber \\
&&\hspace{.15in}+{\cal F}_{1,1;1,1}^0-{\cal F}_{1,0;1,0}^0+{\cal F}_{1,1;1,-1}^0)/24 \nonumber \\
k^\prime&&=({\cal F}_{1,1;1,1}^1-{\cal F}_{1,0;1,0}^1+{\cal F}_{1,1;1,-1}^1 \nonumber \\
&&\hspace{.1in}-{\cal F}_{1,1;1,1}^0+{\cal F}_{1,0;1,0}^0-{\cal F}_{1,1;1,-1}^0)/24 \nonumber \\
\ell            &&=(3{\cal F}_{1,1;0,0}^1+{\cal F}_{1,1;0,0}^0)/4\sqrt{2}\,, \nonumber \\
\ell^\prime&&=({\cal F}_{1,1;0,0}^1-{\cal F}_{1,1;0,0}^0)/4\sqrt{2}\,,
\label{projform}
\end{eqnarray}
with the notation ${\cal F}_{S,m_s;S^\prime,m_s^\prime}^T = 
\langle S m_s T | {\cal F} | S^\prime m_s^\prime T \rangle$. 

The quasiparticle interaction is defined as the second functional derivative of the
energy ${E}$ with respect to the occupation probabilities $n_{{\vec p} s t}$:
\begin{eqnarray}
\delta {E} &&= \sum_{\vec p_1 s_1 t_1} \epsilon_{\vec p_1}\, \delta 
n_{\vec p_1 s_1 t_1} \nonumber \\
&&\hspace{-.2in} + \frac{1}{2\Omega} \sum_{\substack{{\vec p}_1 s_1 t_1
 \\ {\vec p}_2 s_2 t_2}}{\cal F}({\vec p}_1 s_1 t_1;
{\vec p}_2 s_2 t_2) \delta n_{{\vec p}_1 s_1 t_1}
\delta n_{{\vec p}_2 s_2 t_2},
\label{deltae}
\end{eqnarray}
where $\Omega$ is a normalization volume. The quasiparticle 
interaction ${\cal F}$ in momentum space
has units fm$^2$, $s_i$ labels the spin quantum number of quasiparticle $i$,
and $t_i$ labels the isospin quantum number. 
In the present work we consider contributions to the quasiparticle interaction up to 
second order in many-body perturbation theory. 


\subsection{Two-body force contributions}

The first- and second-order 
terms in the perturbative expansion of the ground-state energy from two-body
forces are given by 
\begin{equation}
{E}^{(1)}_{2N} = \frac{1}{2}\sum_{ij} \,
n_i n_j \langle i j \left | \overline{V}_{2N} \right | i j \rangle,
\label{e1}
\end{equation}
\begin{equation}
{E}^{(2)}_{2N} = \frac{1}{4} 
\sum_{ijmn} \left| \langle ij \left | \overline{V}_{2N} \right | mn \rangle \right |^2
\frac{n_i n_j \bar{n}_m \bar{n}_n}
{e_i+e_j-e_m-e_n},
\label{e2}
\end{equation}
where $n_j=\theta(k_f-|\vec k_j|)$, $\bar n_j=\theta(|\vec k_j|-k_f)$, and
$\overline V$ indicates an antisymmetrized interaction.
In Eqs.\ (\ref{e1}) and (\ref{e2}) the sums run over 
momentum, spin and isospin.

\begin{figure*}[t]
\begin{center}
\includegraphics[height=3.5cm]{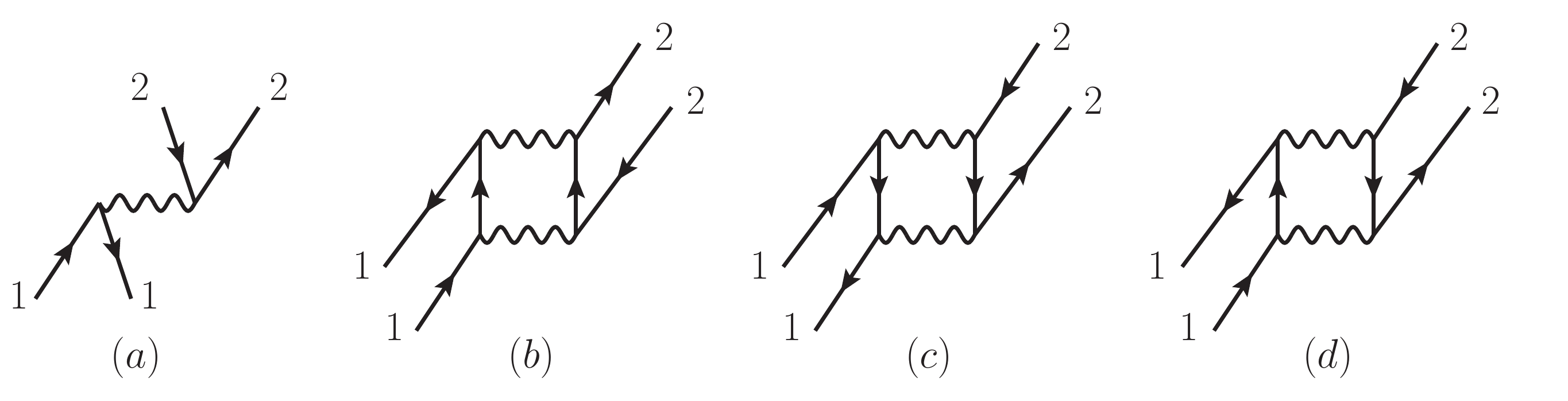}
\end{center}
\vspace{-.5cm}
\caption{Diagrams contributing to the quasiparticle interaction (all 
interactions represented by wavy lines are antisymmetrized): (a) first-order, 
(b) second-order particle-particle, (c) second-order hole-hole, and 
(d) second-order particle-hole diagrams.}
\label{diag}
\end{figure*}

Functionally differentiating Eq.\ (\ref{e1}) with respect to 
$n_1$ and $n_2$ yields for the first-order contribution to the quasiparticle interaction
\begin{eqnarray}
&&\hspace{-.5in}{\cal F}^{(1)}_{2N}({\vec p}_1 s_1 t_1; {\vec p}_2 s_2 t_2) 
= \langle 12 | \overline V_{2N} | 12 \rangle \nonumber \\ 
&& = \langle {\vec p}_1 s_1 t_1; {\vec p}_2 s_2 t_2 | 
\overline V_{2N} | {\vec p}_1 s_1 t_1 ; {\vec p}_2 s_2 t_2 \rangle.
\label{qpi1}
\end{eqnarray}
shown diagrammatically in Fig.\ \ref{diag}(a).
Since this is just the antisymmetrized free-space nucleon-nucleon potential,
only the four central and two exchange-tensor terms in the quasiparticle interaction 
can be generated and the total spin $\vec S = (\vec \sigma_1+ \vec \sigma_2)/2$ 
(with associated quantum number $S$) is conserved.
From the second-order energy in Eq.\ (\ref{e2}), three different 
contributions to the quasiparticle interaction arise which are distinguished by intermediate 
particle-particle, hole-hole, and particle-hole states shown diagrammatically
in Figs.\ \ref{diag}(b), \ref{diag}(c), and \ref{diag}(d), respectively. They have the form
\begin{equation}
{\cal F}^{(2pp)}_{2N} ({\vec p}_1 s_1 t_1; {\vec p}_2 s_2 t_2) 
= \frac{1}{2} \sum_{mn} \frac{|\langle 12 | \overline V_{2N} | mn \rangle|^2 \bar n_m
\bar n_n} {\epsilon_1 + \epsilon_2 - \epsilon_m - \epsilon_n}
\label{qpi2pp}
\end{equation}
\begin{equation}
{\cal F}^{(2hh)}_{2N} ({\vec p}_1 s_1 t_1; {\vec p}_2 s_2 t_2)
=\frac{1}{2} \sum_{ij} \frac{|\langle ij | \overline V_{2N} | 12 \rangle |^2 n_i n_j}
{\epsilon_i + \epsilon_j - \epsilon_1 - \epsilon_2}
\label{qpi2hh}
\end{equation}
\begin{equation}
{\cal F}^{(2ph)}_{2N} ({\vec p}_1 s_1 t_1; {\vec p}_2 s_2 t_2) 
= -2 \sum_{jn} \frac{|\langle 1j | \overline V_{2N} | 2n \rangle |^2 n_j \bar n_n}
{\epsilon_1 + \epsilon_j - \epsilon_2 - \epsilon_n}.
\label{qpi2ph}
\end{equation}

Eqs.\ (\ref{qpi1})--(\ref{qpi2ph}) can be evaluated for realistic nuclear interactions by
first decomposing the potential matrix elements into a partial-wave sum. The Fermi liquid
parameters are then obtained by integrating over the angle $\theta$ between $\vec p_1$
and $\vec p_2$ with appropriate Legendre polynomials as weighting functions.
For the first-order term, as well as the second-order particle-particle and
hole-hole diagrams, the partial-wave decomposition is 
straightforward since the two quasiparticle states are both in the incoming or outgoing state. 
However, the evaluation of the second-order particle-hole diagram is more complicated due to 
the cross-coupling of the quasiparticle states in one incoming and one outgoing state. We state
here the final expressions, and for additional details the reader is referred to Ref.\ \cite{holt13prca}.
As already mentioned, the first-order contribution to the quasiparticle interaction
is just the antisymmetrized free-space potential:
\begin{widetext}
\begin{eqnarray}
{\cal F}^{(1)}_L (S m_s m_s^\prime;T)&=& 16 \pi (2L+1) \sum_{l l^\prime J}
i^{l- l^\prime} \sqrt{(2l+1)(2 l^\prime +1)} 
\, \langle l 0 \, S m_s | J M \rangle \nonumber \\ 
&&\times  \langle l^\prime 0 \, S m_s^\prime
| JM \rangle \int_0^{k_f} dp\, \frac{p}{k_f^2}  \, \langle plSJT | V_{2N} | 
pl^\prime SJT \rangle P_L(1-2p^2/k_f^2),
\label{qp1}
\end{eqnarray}
\end{widetext}
where $p=q/2$. The second-order terms are given by
\begin{widetext}
\begin{eqnarray}
&&{\cal F}^{(2pp)}_L(S m_s m_s^\prime; T) = \frac{16(2L+1)}{k_F^2} 
\sum_{\substack {l_1l_2l_3l_4mm^\prime \\ 
\bar m \bar m_s JJ^\prime M}} \int_0^{k_F} dp \, p P_L(1-2p^2/k_f^2) \int_p^{\infty} dk \, k^2 
N(l_1ml_2\bar m l_3m^\prime l_4 \bar m) 
P_{l_1}^m(0) P_{l_3}^{m^\prime}(0) \nonumber \\
&& \hspace{.3in} \times   \frac{M}{p^2-k^2} i^{l_2+l_3-l_1-l_4} {\cal C}^{JM}_{l_1m S m_s} 
{\cal C}^{JM}_{l_2\bar m S \bar m_s} 
{\cal C}^{J^\prime M}_{l_3 m^\prime S m_s^\prime} {\cal C}^{J^\prime M}_{l_4 \bar m S \bar m_s}  
\int^{{\rm min}\{ x_0,1 \}}_{{\rm max}\{ -x_0,-1 \}} 
d x \, P_{l_2}^{\bar m}(x) P_{l_4}^{\bar m}(x) \nonumber \\
&& \hspace{.3in} \times  \langle pl_1SJT | V_{2N} | kl_2SJT \rangle
\langle kl_4SJ^\prime T | V_{2N} | pl_3SJ^\prime T \rangle,
\label{pp2}
\end{eqnarray}

\begin{eqnarray}
&&{\cal F}^{(2hh)}_L(S m_s m_s^\prime; T) = \frac{16(2L+1)}{k_F^2} 
\sum_{\substack {l_1l_2l_3l_4mm^\prime \\ 
\bar m \bar m_s JJ^\prime M}} \int_0^{k_F} dp \, p P_L(1-2p^2/k_f^2) \int_0^p dk \, k^2 
N(l_1ml_2\bar m l_3m^\prime l_4 \bar m) 
P_{l_1}^m(0) P_{l_3}^{m^\prime}(0) \nonumber \\
&& \hspace{.3in} \times   \frac{M}{k^2-p^2} i^{l_2+l_3-l_1-l_4} {\cal C}^{JM}_{l_1m S m_s} 
{\cal C}^{JM}_{l_2\bar m S \bar m_s} 
{\cal C}^{J^\prime M}_{l_3 m^\prime S m_s^\prime} {\cal C}^{J^\prime M}_{l_4 \bar m S \bar m_s}  
\int^{{\rm min}\{ -x_0,1 \}}_{{\rm max}\{ x_0,-1 \}} 
d x \, P_{l_2}^{\bar m}(x) P_{l_4}^{\bar m}(x) \nonumber \\
&& \hspace{.3in} \times  \langle pl_1SJT | V_{2N} | kl_2SJT \rangle
\langle kl_4SJ^\prime T | V_{2N} | pl_3SJ^\prime T \rangle,
\label{hh2}
\end{eqnarray}

\begin{eqnarray}
{\cal F}_L^{(2ph)}(s_1 s_2 s_1^\prime s_2^\prime;t_1 t_2 t_1^\prime t_2^\prime) 
&=& \frac{16(2L+1)}{\pi k_f^2} \int_{0}^{k_f} dp \, p P_L(1-2p^2/k_f^2) 
\int_0^{2\pi} d \phi_3 \int_{{\rm max}\{0, y_0\}}^{k_f}
dk_3  k_3^2  \int_{{\rm max}\{-1,z_0 \}}^1 d \cos \theta_3  
 \nonumber \\ 
&& \hspace{-.9in} \times \sum_{\substack {l_1l_2 l_3 l_4 s_3 s_4 \\ 
m_1 m_2 m_3 m_4}} 
\langle p^\prime l_1 m_1 s_1 s_3 t_1 t_3 | V | k^\prime l_2 m_2 s_2 s_4 t_2 t_4 \rangle 
\langle k^\prime l_4 m_4 s_2^\prime s_4 t_2^\prime t_4 | V | p^\prime l_3 m_3 s_1^\prime 
s_3 t_1^\prime t_3 \rangle \nonumber \\
&& \hspace{-.9in} \times \cos ((m_3 - m_1 + m_2 - m_4)\phi_{p^\prime})
P_{l_1}^{m_1}(\cos \theta_{p^\prime}) P_{l_2}^{m_2}(\cos \theta_{k^\prime})
P_{l_3}^{m_3}(\cos \theta_{p^\prime}) P_{l_4}^{m_4}(\cos \theta_{k^\prime}) \nonumber \\
&& \hspace{-.9in} \times i^{l_2+l_3-l_1-l_4}  N(l_1 m_1 l_2 m_2 l_3 m_3 l_4 m_4) 
\frac{M}{p^2 + k_3 p \cos \theta_3},
\label{ph2}
\end{eqnarray}
\end{widetext}
where in the particle-particle and hole-hole diagrams $\vec k = (\vec k_3 - \vec k_4)/2$, 
$x_0 = (k^2-p^2)/(2k\sqrt{k_f^2-p^2})$, $x=\cos \theta_k$, $P_l^m$ are the associated 
Legendre functions, and $N(l_1 m_1 l_2 m_2 l_3 m_3 l_4 m_4) = N_{l_1}^{m_1}
N_{l_2}^{m_2} N_{l_3}^{m_3} N_{l_4}^{m_4}$ with $N_l^m = \sqrt{(2l+1)(l-m)!/(l+m)!}$.
In the particle-hole diagram we have additionally $\vec p^{\, \prime} = (\vec p_1 - \vec k_3)/2$, 
$\vec k^\prime = (\vec p_2 - \vec k_4)/2$, $y_0 = k_f-2p$, 
and $z_0 = (k_f^2-k_3^2-4p^2)/(4 k_3 p)$.
From the matrix elements of the second-order particle-hole contribution 
in the uncoupled spin and isospin basis, it is trivial through recoupling to generate
the terms needed in Eq.\ (\ref{projform}) to extract the Fermi liquid parameters.

\begin{figure*}
\begin{center}
\minipage{0.325\textwidth}
\includegraphics[scale=0.46,clip]{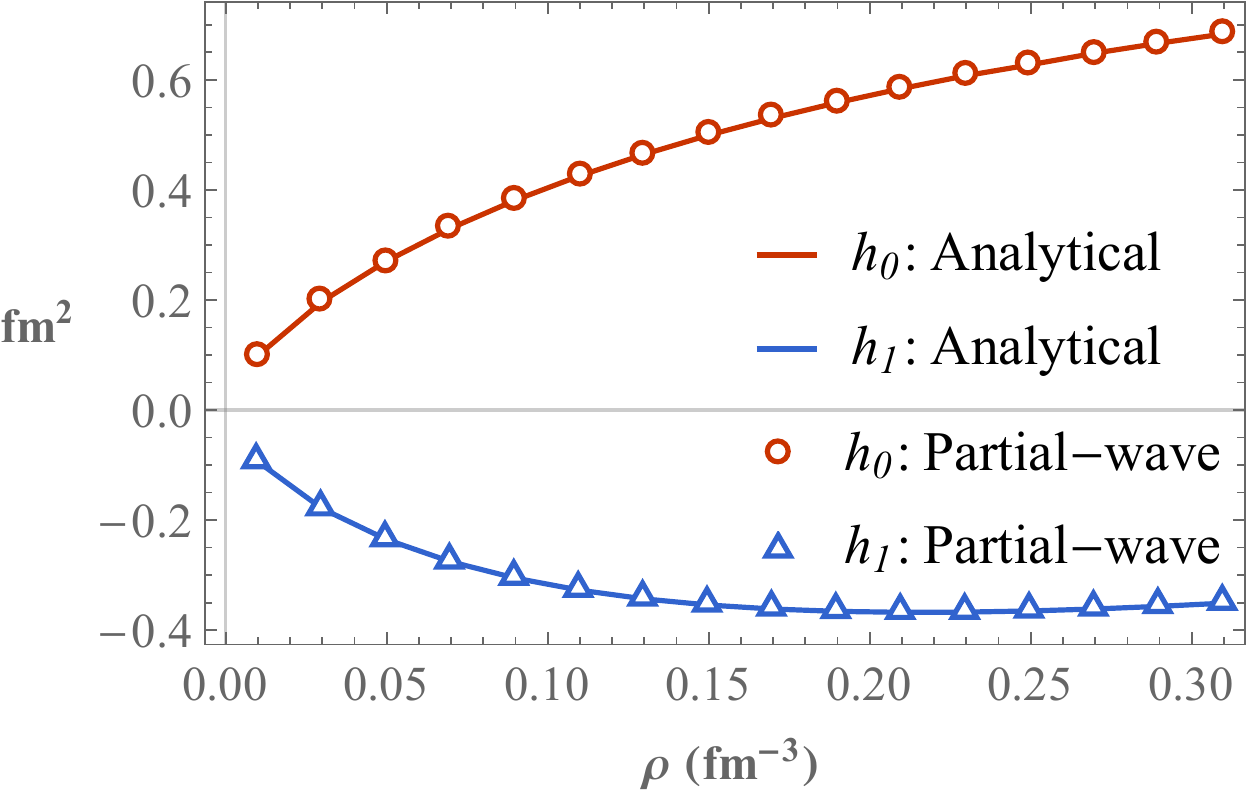}
\endminipage \hfill
\minipage{0.325\textwidth}
\includegraphics[scale=0.46,clip]{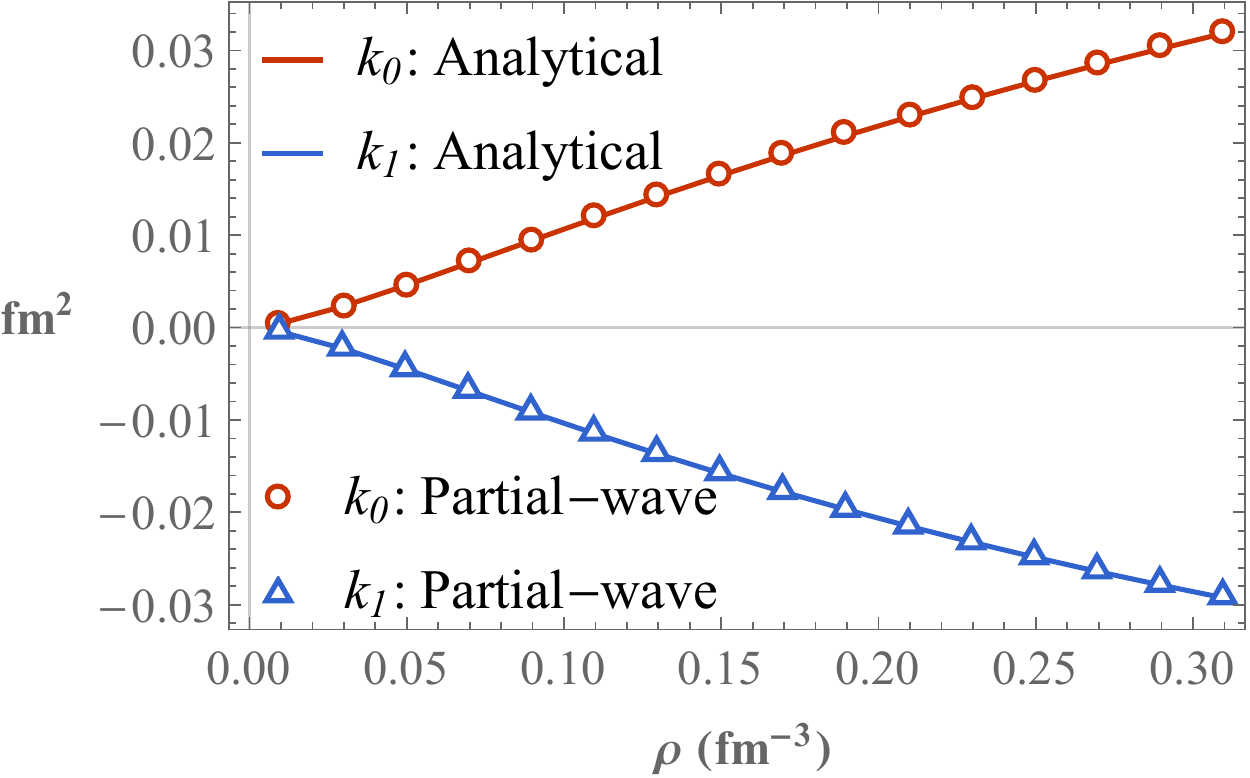}
\endminipage \hfill
\minipage{0.325\textwidth}
\includegraphics[scale=0.46,clip]{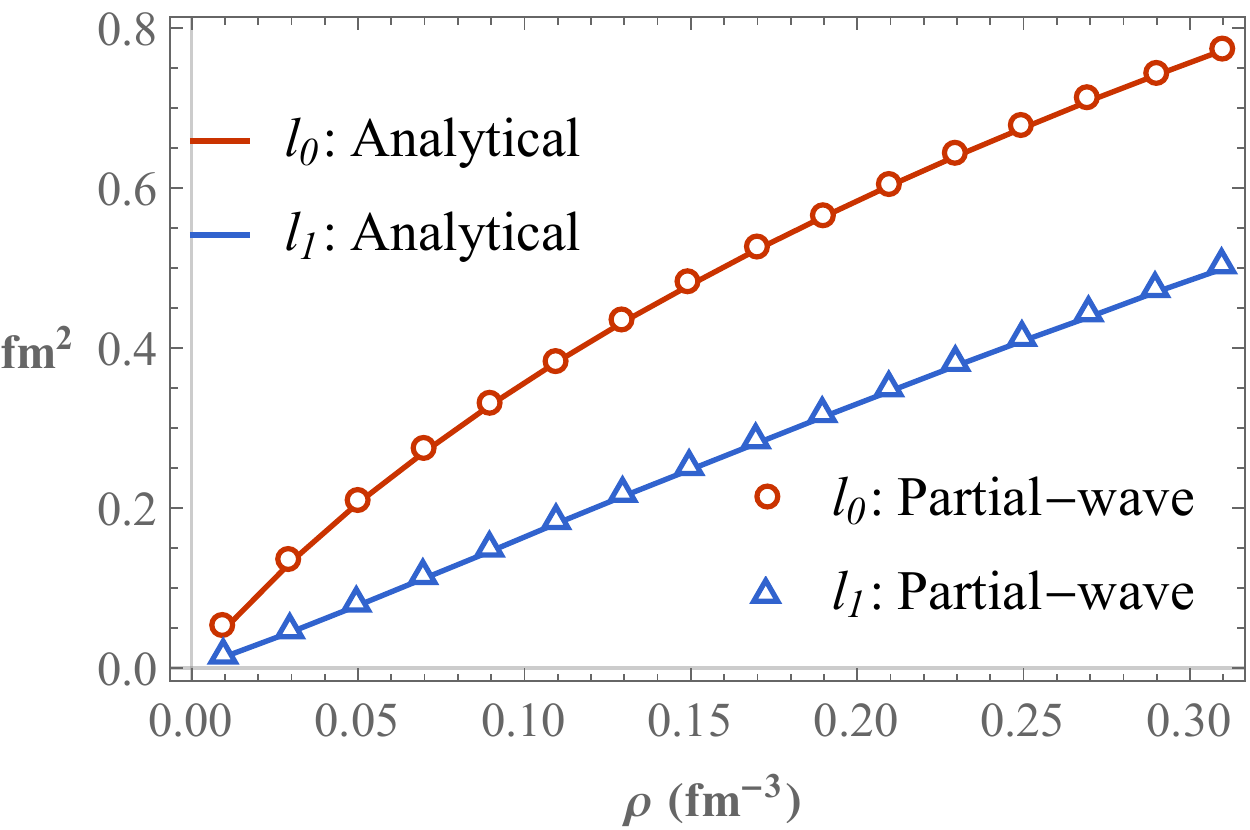}
\endminipage \hfill
\caption{Comparison of the $L=0,1$ noncentral Fermi liquid parameters in
nuclear matter from modified pseudoscalar exchange at second order (left and
middle plots) as well as from the interference of a central and spin-orbit
interaction in the particle-hole channel (right plot). Results from both a partial-wave
decomposition and semi-analytical calculation are compared and found to be
in excellent agreement.}
\label{testing}
\end{center}
\end{figure*}

Given the numerical complexity of Eqs.\ (\ref{pp2})--(\ref{ph2}), we have benchmarked our
codes against semi-analytical results from model interactions. As a first case we consider a 
modified pseudoscalar interaction of the form
\begin{equation} 
V_{ps}= g^2 \frac{\vec \sigma_1 \cdot \vec q\,\,\vec \sigma_2 \cdot 
\vec q}{(m^2+q^2)^2} \vec \tau_1 \cdot \vec \tau_2 \, ,
\label{mopem}
\end{equation}
where $\vec q$ is the momentum transfer, $g$ is a dimensionless coupling constant,
and $m$ is the mass parameter chosen to be large enough to achieve good convergence 
in momentum integrals and partial wave
summations. We choose for concreteness $g=6$ and $m=600$\,MeV.
As a second interesting case, we consider the interference 
between an isoscalar central and spin-orbit interaction of the form
\begin{equation} 
V_c = g^2 \frac{1}{m^2+q^2} \, \,,
\label{cent}
\end{equation}
\begin{equation} 
V_{so} = \frac{g^2}{m^2} \frac{i \, (\vec \sigma_1 +\vec \sigma_2) \cdot 
(\vec q \times \vec p\,)}{m^2+q^2} \, \,,
\label{soi}
\end{equation}
where $\vec p$ is the incoming relative momentum. We choose the same values for $g$
and $m$ as in the modified pseudoscalar case above. 

In Ref.\ \cite{holt13prca} it was shown that the second-order particle-particle 
and hole-hole diagrams can only generate the central, relative momentum tensor, and 
center-of-mass tensor components of the quasiparticle interaction. In fact, only the 
combination of a spin-orbit force with any non-spin-orbit force in the second-order particle-hole 
diagram can lead to a cross-vector term. These general conclusions are exemplified in the
test interactions considered in Eqs.\ (\ref{mopem})--(\ref{soi}). 
In particular, the $L=0,1$ isoscalar relative-momentum tensor and 
center-of-mass tensor Fermi liquid parameters from modified pseudoscalar exchange at
second order are shown in the left and middle plots of 
Fig.\ \ref{testing} employing both the partial-wave decomposition in 
Eqs.\ (\ref{pp2})--(\ref{ph2}) as well as semi-analytical expressions similar to those 
in Ref.\ \cite{holt13prca}. 
The isovector contributions only differ from the isoscalar contributions by integer factors and 
therefore are not shown explicitly. The modified pseudoscalar interaction at second order 
also gives rise to central components of the quasiparticle interaction, but these
have been considered in previous work \cite{holt11npa}.
From Fig.\ \ref{testing} we see that the numerical agreement across
the full range of densities considered, $0 < \rho < 2 \rho_0$, is excellent.
In the rightmost plot of Fig.\ \ref{testing} we show the Fermi
liquid parameters associated with the cross-vector interaction from the interference term between
a central and spin-orbit force. Again the numerical agreement between the two methods
is very good.


\subsection{Three-body force contributions}

We next consider contributions to the quasiparticle interaction from three-body forces.
The Hartree-Fock energy is given by
\begin{equation}
{E}^{(1)}_{3N} = \frac{1}{6}\sum_{ijk}
n_i n_j n_k \langle ijk | \overline{V}_{3N} | ijk \rangle,
\label{e3nf}
\end{equation}
where the totally antisymmetrized three-body potential is given by $\overline V_{3N} = 
( 1 - P_{12} - P_{13} - P_{23} + P_{12}P_{23} + P_{12}P_{13} )V_{3N}$.
In the present work we employ the N2LO chiral three-body force, which includes a long-range
two-pion exchange component $V_{3N}^{2\pi}$, a one-pion exchange contribution $V_{3N}^{1\pi}$, 
and a pure contact force $V_{3N}^{\rm cont}$. The two-pion exchange three-nucleon interaction 
has the form
\begin{equation}
V_{3N}^{2\pi} = \sum_{i\neq j\neq k} \frac{g_A^2}{8f_\pi^4} 
\frac{\vec{\sigma}_i \cdot \vec{q}_i \, \vec{\sigma}_j \cdot
\vec{q}_j}{(\vec{q_i}^2 + m_\pi^2)(\vec{q_j}^2+m_\pi^2)}
F_{ijk}^{\alpha \beta}\tau_i^\alpha \tau_j^\beta,
\label{3n1}
\end{equation}
where $g_A=1.29$, $f_\pi = 92.2$ MeV, $m_{\pi} = 138$ MeV is the average pion mass, 
$\vec{q}_i$ denotes the difference between the final and initial momenta of nucleon $i$, and 
the isospin tensor $F_{ijk}^{\alpha \beta}$ is given by
\begin{equation}
F_{ijk}^{\alpha \beta} = \delta^{\alpha \beta}\left (-4c_1m_\pi^2
 + 2c_3 \vec{q}_i \cdot \vec{q}_j \right ) + 
c_4 \epsilon^{\alpha \beta \gamma} \tau_k^\gamma \vec{\sigma}_k
\cdot \left ( \vec{q}_i \times \vec{q}_j \right ).
\label{3n4}
\end{equation}
The one-pion exchange component of the three-nucleon interaction is
defined by
\begin{equation}
V_{3N}^{1\pi} = -\sum_{i\neq j\neq k} \frac{g_A c_D}{8f_\pi^4 \Lambda_\chi} 
\frac{\vec{\sigma}_j \cdot \vec{q}_j}{\vec{q_j}^2+m_\pi^2} \vec{\sigma}_i \cdot
\vec{q}_j \, {\vec \tau}_i \cdot {\vec \tau}_j \, ,
\label{3n2}
\end{equation}
where $\Lambda_{\chi} = 700$ MeV. 
Finally, the three-nucleon contact interaction has the form
\begin{equation}
V_{3N}^{\rm cont} = \sum_{i\neq j\neq k} \frac{c_E}{2f_\pi^4 \Lambda_\chi}
{\vec \tau}_i \cdot {\vec \tau}_j\, .
\label{3n3}
\end{equation}

\begin{figure*}
\begin{center}
\includegraphics[scale=0.55,clip]{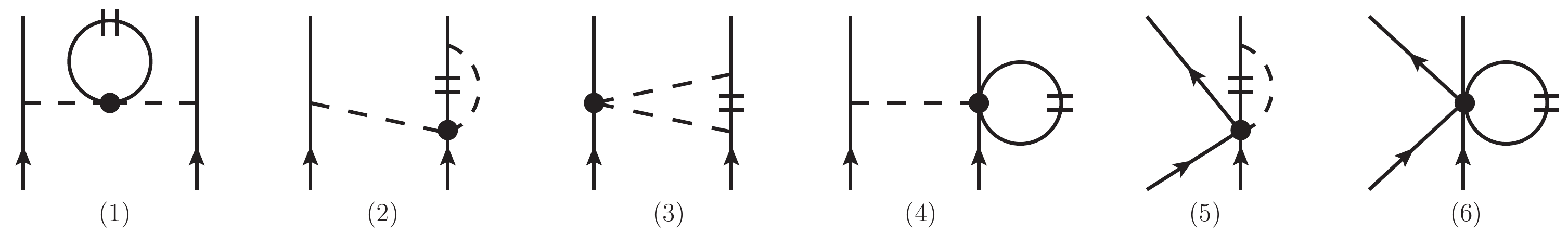}
\end{center}
\vspace{-.5cm}
\caption{Diagrammatic contributions to the quasiparticle interaction in symmetric nuclear
matter generated from the three terms in the N2LO chiral three-body force. 
The short double-line symbolizes summation over the filled Fermi sea of nucleons. 
Reflected diagrams of (2) and (3) are not shown.}
\label{mednn}
\end{figure*}

In pure neutron matter only the terms proportional to $c_1$ and $c_3$ contribute to the
ground state energy and quasiparticle interaction, but for symmetric nuclear matter in general
all terms are needed. 
Taking two functional derivatives of Eq.\ (\ref{e3nf}) with respect to $n_1$ and $n_2$ yields
\begin{eqnarray}
&&\hspace{-.4in} {\cal F}^{(1)}_{3N}({\vec p}_1 s_1 t_1 , {\vec p}_2 s_2 t_2) = \frac{1}{3} \sum_i
 n_i \left [ \langle i12 | \overline{V}_{3N} | i12 \rangle \right . \nonumber \\
 && \hspace{.5in} \left . + \langle 1i2 | \overline{V}_{3N} | 1i2 \rangle 
 + \langle 12i | \overline{V}_{3N} | 12i \rangle\right ].
\label{qpi3na}
\end{eqnarray}
Since the three-body force is symmetric under the interchange of particle labels, we can rewrite
Eq.\ (\ref{qpi3na}) without loss of generality as
\be
\hspace{-.4in} {\cal F}^{(1)}_{3N}({\vec p}_1 s_1 t_1 , {\vec p}_2 s_2 t_2) = \sum_i
 n_i \langle i12 | \overline{V}_{3N} | i12 \rangle.
\label{qpi3nb}
\ee
In general there are nine distinct direct (and exchange) contributions to the quasiparticle interaction 
from a three-body force. In Fig.\ \ref{mednn} we show the direct terms from the N2LO chiral
three-nucleon interaction (exchange terms can be obtained by swapping the two outgoing lines). 
As seen in Fig.\ \ref{mednn} there are three topologically distinct contributions from the two-pion exchange 
three-body force $V_{3N}^{2\pi}$. 
Contribution `(2)' represents the sum of four reflected diagrams, while contribution 
`(3)' represents the sum of two reflected diagrams. The one-pion
exchange contribution $V_{3N}^{1\pi}$ gives rise to two topologically distinct diagrams, shown as 
`(4)' and `(5)' in Fig.\ \ref{mednn}. Finally, there is a single diagram `(6)' coming from the three-body
contact force $V_{3N}^{\rm cont}$. As shown in the Appendix, this diagram
contributes only to the central components of the quasiparticle interaction.

At second order in perturbation theory we include the effects of three-body forces by first 
constructing a density-dependent two-body force $V_{2N}^{\rm med}$, as described in detail 
in Refs.\ \cite{holt09,holt10}. In Eqs.\ (\ref{qpi2pp})--(\ref{qpi2ph}) we then replace $\overline V_{2N}$ 
with $\overline V_{2N}^{\rm eff} = \overline V_{2N} + \overline V_{2N}^{\rm med}$. This approximation
accounts for only a subset of the full second-order contributions from three-body forces.


\section{Results}
\label{results}

In the present section we focus on the noncentral components of the quasiparticle interaction 
from the five different chiral nuclear forces \{n2lo450, n2lo500, n3lo414, n3lo450, n3lo500\}
\cite{entem03,coraggio07,marji13,coraggio13,coraggio14,sammarruca15}. 
We focus primarily on the role of three-body forces and second-order perturbative contributions.
The quality of the nuclear force models and perturbative many-body method is benchmarked
by comparing the nuclear incompressibility, isospin asymmetry energy, and effective mass (which
are related to specific central Fermi liquid parameters) with empirical values. 
We also study the convergence of the Legendre polynomial decomposition for both central
and noncentral forces.


\subsection{First-order perturbative contributions to Fermi liquid parameters}
\label{lp3nf}

\begin{figure*}[t]
\begin{center}
\minipage{0.49\textwidth}
\includegraphics[scale=0.9,clip]{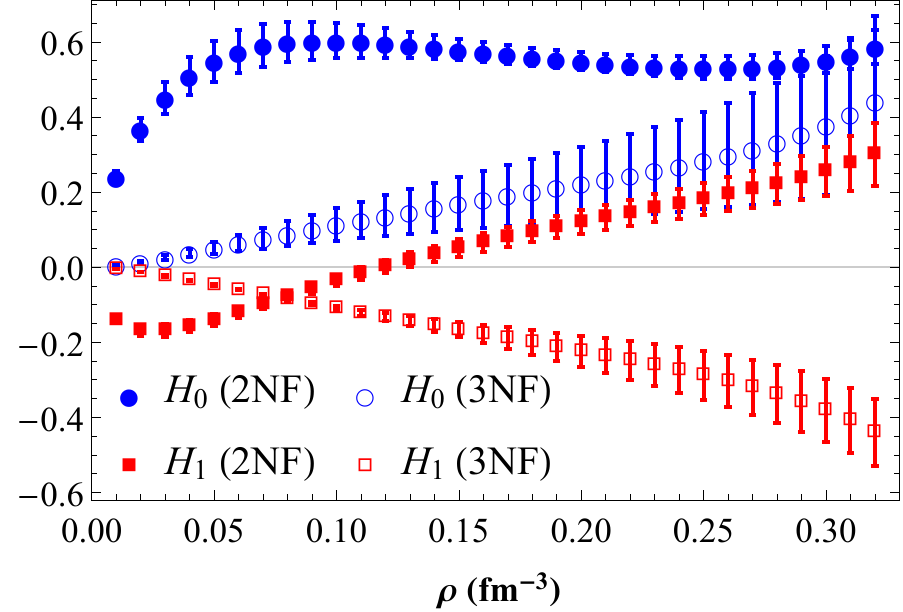}
\endminipage \hfill
\minipage{0.49\textwidth}
\includegraphics[scale=0.9,clip]{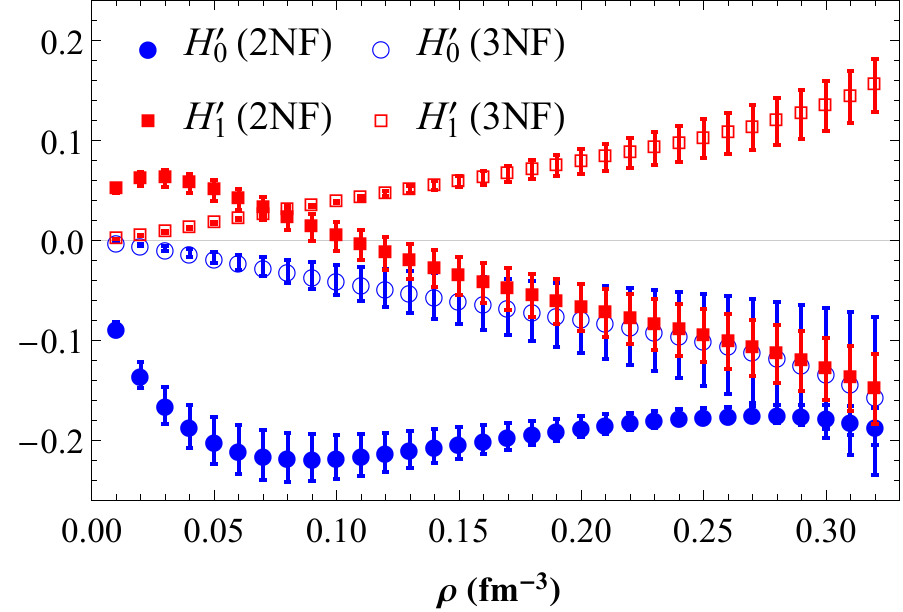}
\endminipage \hfill
\caption{First-order perturbative contributions to the dimensionless $L=0,1$ Fermi
liquid parameters of the relative tensor interactions as a 
function of density. Results for both two- and three-body forces are included.}
\label{hfh}
\end{center}
\end{figure*}

At first order in perturbation theory, two-nucleon forces contribute only to the relative momentum
tensor noncentral Fermi liquid parameters ($H_L, H_L^\prime$) due to the underlying Galilean invariance 
of the free-space interaction. In Fig.\ \ref{hfh} we show as solid-circle and solid-square 
symbols the magnitude of the dimensionless Fermi liquid parameters ($H_0, H_1$) and 
($H_0^\prime, H_1^\prime$) from two-body forces as a function of density. The 
error bars are calculated as the standard deviation of the results from the five nucleon-nucleon 
potentials considered in the present work and represent an estimate of systematic uncertainties 
in the construction of realistic two- and three-body force models.
Note that in this section, we employ the Landau effective mass that enters into the 
density of states, Eq.\ (\ref{dos}), computed from the full quasiparticle interaction including
second-order perturbative contributions (see Eq.\ (\ref{effmass}) below). 
Comparing the results from two-body forces in Fig.\ \ref{hfh}, we find the approximate 
relationship $H_0 \simeq -3 H_0^\prime$ and $H_1 \simeq -3 H_1^\prime$, which in fact 
exactly holds for all $L$ in the case of a pure one-pion exchange (OPE) nucleon-nucleon 
potential \cite{brown77}. 

Next we present analytical expressions for the noncentral Fermi-liquid parameters in nuclear matter
from the N2LO chiral three-nucleon interaction. The associated low-energy constants $\{ c_1, 
c_3, c_4, c_D, c_E\}$ have been fitted separately
for each nuclear potential and are compiled in Table \ref{lectab}. 
\begin{table}[h]
\begin{tabular}{|c|c|c|c|c|c|}\hline
& $c_1$ & $c_3$ & $c_4$ & $c_D$ & $c_E$ \\ \hline
n2lo450 & $-0.81$ & $-3.40$ & $3.40$ & $-0.326$ & $-0.149$ \\ \hline
n2lo500 & $-0.81$ & $-3.40$ & $3.40$ & $-0.165$ & $-0.169$ \\ \hline
n3lo414 & $-0.81$ & $-3.00$ & $3.40$ & $-0.400$ & $-0.072$ \\ \hline
n3lo450 & $-0.81$ & $-3.40$ & $3.40$ & $-0.240$ & $-0.106$ \\ \hline
n3lo500 & $-0.81$ & $-3.20$ & $5.40$ & $-0.200$ & $-0.205$ \\ \hline
\end{tabular}
\caption{Low-energy constants associated with the N2LO chiral three-body force for
the five different nuclear potentials considered in the present work. The constants 
$c_1$, $c_3$, and $c_4$ have units GeV$^{-1}$, while $c_D$ and $c_E$ are
unitless.}
\label{lectab}
\end{table}
We present individually the Fermi liquid parameters arising from the five 
diagrammatic contributions in
Fig.\ \ref{mednn}. The pion self-energy correction $V_{NN}^{\rm med,1}$
leads to a relative-momentum tensor interaction

\begin{eqnarray} 
&& \hspace{-.5in} h_L=-3h'_L = {g_A^2 m_\pi^3 u^5\over 3\pi^2 f_\pi^4}
\int_{-1}^1 \!dz\,(1-z) (2L+1) \nonumber \\ 
&& \hspace{.5in} \times P_L(z)\,{c_3 u^2(z-1)-c_1\over 
[1+2u^2(1-z)]^2}\,,
\end{eqnarray}
where $u = k_f/m_\pi$ and $P_L(z)$ is the Legendre polynomial of degree $L$.
The pion-exchange vertex correction, $V_{NN}^{\rm med,2}$,
 likewise gives rise to relative-tensor force of the form
\begin{eqnarray}
&& \hspace{-.12in} h_L=-3h'_L ={g_A^2 m_\pi^3\over 
\pi^2(4f_\pi)^4}\int_{-1}^1 \!dz\,
{(1-z) (2L+1)P_L(z) \over 1+2u^2(1-z)} \nonumber \\
&& \times \bigg\{{64u^2\over 3}(c_3-2c_4) \arctan 2u +\bigg[ {(c_3+c_4)({1\over 3}-z)-4c_1\over u} 
\nonumber \\ && +4u\big(3c_4-4c_1-c_3-(c_3+c_4)z\big)+\bigg] \ln(1+4u^2)\nonumber \\ &&  +4(c_3+c_4)u z
\bigg(1+2u^2-{8u^4\over 3}\bigg)+{32u^5\over 9}(5c_3-7c_4)\nonumber \\ 
&& + 8u^3
(4c_1-3c_3+5c_4)+{4u\over 3}(12c_1-c_3-c_4)\bigg\} \,.
\end{eqnarray}
Since $V_{NN}^{\rm med,1}$ and $V_{NN}^{\rm med,2}$ renormalize the one-pion exchange 
interaction through self-energy and vertex corrections, the associated Fermi liquid parameters
obey the generic relationship $h_L = -3h_L^\prime$ as seen explicitly above.

The Pauli-blocked two-pion exchange contribution, $V_{NN}^{\rm med,3}$, gives rise to a
richer set of spin and isospin structures that lead to contributions to all of the noncentral Fermi liquid
parameters ($h,h^\prime, k,k^\prime, \ell,\ell^\prime$). For the relative tensor interaction
we find
\begin{eqnarray}
&& \hspace{-.1in} h_L = -3h'_L \\ \nonumber
&& \hspace{-.05in} ={g_A^2 c_4 m_\pi^3 u^2\over 16\pi^3 f_\pi^4}
\int_0^u\!dl\,l^2 \int_{-1}^1\!dx \int_{-1}^1\!dy \int_0^\pi\!d\phi\, 
(2L+1)P_L(z) \\ \nonumber 
&& \hspace{-.05in} \times {(1-z)[u(1+z)-l(x+y)]^2\over (1+z)(1+u^2+l^2-2u l x)
(1+u^2+l^2-2u l y)}\,, 
\end{eqnarray}
with $z = xy +\sqrt{(1-x^2)(1-y^2)}\cos\phi$. The center-of-mass tensor Fermi liquid
parameters in the isoscalar and isovector channel are given by

\begin{widetext}
\begin{eqnarray}k_L=-3k'_L &=&{g_A^2 c_4 m_\pi^3 u^2\over 16\pi^3 f_\pi^4}
\int_0^u\!dl\,l^2 \int_{-1}^1\!dx \int_{-1}^1\!dy \int_0^\pi\!d\phi\, 
(2L+1)P_L(z)
\nonumber \\ && \times {(1-z)[u(1+z)-l(x+y)]^2+2l^2(x^2+y^2+z^2-1-2x y z)
\over (1+z)(1+u^2+l^2-2u l x)(1+u^2+l^2-2u l y)}\,,
\end{eqnarray}
\end{widetext}

We note that the relative tensor interaction exhibits the same relationship, 
$k_L=-3k'_L$, between the isoscalar and isovector components as the 
relative-momentum tensor interactions above. Finally, the Pauli-blocked two-pion
exchange diagram produces a cross-vector interaction with Fermi liquid parameters

\begin{widetext}
\begin{eqnarray}
\ell_L &=&{3g_A^2 m_\pi^3 u\over 32\pi^3 
f_\pi^4}\int_0^u\!dl\,l^2
\int_{-1}^1\!dx \int_{-1}^1\!dy \int_0^\pi\!d\phi\, 
(2L+1)P_L(z)\sqrt{{1-z\over
1+z}}\nonumber \\ && \times[u(1+z)-l(x+y)] {2c_1+(c_3-c_4)[l^2-u 
l(x+y)+u^2z]
\over (1+u^2+l^2-2u l x)(1+u^2+l^2-2u l y)}\,,
\label{ell1}
\end{eqnarray}

\begin{eqnarray}
\ell'_L &=&{g_A^2 m_\pi^3 u\over 32\pi^3 
f_\pi^4}\int_0^u\!dl\,l^2
\int_{-1}^1\!dx \int_{-1}^1\!dy \int_0^\pi\!d\phi\, 
(2L+1)P_L(z)\sqrt{{1-z\over
1+z}}\nonumber \\ && \times[u(1+z)-l(x+y)] {6c_1+(3c_3+c_4)[l^2-u 
l(x+y)+u^2z]
\over (1+u^2+l^2-2u l x)(1+u^2+l^2-2u l y)}\,,
\end{eqnarray}

\begin{eqnarray}
\tilde \ell_L &=&{3g_A^2 m_\pi^3 u\over 16\pi^3 
f_\pi^4}\int_0^u\!
dl\,l^2 \int_{-1}^1\!dx \int_{-1}^1\!dy \int_0^\pi\!d\phi\, 
(2L+1)P_L(z)\nonumber
\\ && \times[u(1+z)-l(x+y)]{2c_1+(c_3-c_4)[l^2-u l(x+y)+u^2z]
\over (1+u^2+l^2-2u l x)(1+u^2+l^2-2u l y)}\,,
\end{eqnarray}

\begin{eqnarray}
\tilde\ell'_L &=&{g_A^2 m_\pi^3 u\over 16\pi^3 
f_\pi^4}\int_0^u\!
dl\,l^2 \int_{-1}^1\!dx \int_{-1}^1\!dy \int_0^\pi\!d\phi\, 
(2L+1)P_L(z)\nonumber
\\ && \times[u(1+z)-l(x+y)] {6c_1+(3c_3+c_4)[l^2-u l(x+y)+u^2z]
\over (1+u^2+l^2-2u l x)(1+u^2+l^2-2u l y)}\,.
\label{ell4}
\end{eqnarray}
\end{widetext}

In Eqs.\ (\ref{ell1})--(\ref{ell4}) we have employed two parametrizations of the cross-vector 
quasiparticle interaction in terms of Fermi liquid parameters:
\begin{equation}
{\cal F}_{\rm cross} = (\vec \sigma_1 \times \vec \sigma_2)
\cdot (\hat q \times \hat P) \sum_{L=0}^\infty
\ell_L(k_f) \, P_L(\hat p_1 \cdot \hat p_2) \,,
\label{cvn}
\end{equation}
and
\begin{equation} 
{\cal F}_{\rm cross} = {(\vec \sigma_1 \times \vec
\sigma_2) \cdot (\vec p_1 \times \vec p_2) \over |\vec p_1+\vec p_2|^2}
\sum_{L=0}^\infty
\tilde \ell_L(k_f) \, P_L(\hat p_1 \cdot \hat p_2) \,,
\end{equation}
the latter being more convenient in calculations of nuclear response functions 
\cite{davesne15prc}. In pure neutron matter $V_{NN}^{\rm med,3}$ gives rise
to only central and cross-vector interactions \cite{holt13prca}. In the present
symmetric nuclear matter calculation also the relative tensor and center-of-mass
tensor terms are generated due to the three-body force proportional to the 
low-energy constant $c_4$.

The medium-dependent vertex correction, $V_{NN}^{\rm med,4}$, from the 
one-pion exchange three-body force leads to a relative tensor force with 
associated Fermi liquid parameters
\begin{equation} 
h_L=-3h'_L = {g_A c_D m_\pi^3 u^5\over 24\pi^2 
f_\pi^4\Lambda_\chi}
\int_{-1}^1 \!dz\,{(1-z) (2L+1)P_L(z)\over 1+2u^2(1-z)}.
\end{equation}
The second contribution, $V_{NN}^{\rm med,5}$, from the one-pion exchange 
three-body force has the form of a zero-range interaction with vertex 
correction, leading to a finite-range force that contributes to only the 
$L=0,1$ Fermi liquid parameters:
\begin{eqnarray} 
&& \hspace{-.14in} h_0=-h'_0=-h_1= h'_1 = k_0= -k'_0 = k_1= 
-k'_1 \\ \nonumber
&& \hspace{-.05in} ={g_A c_D m_\pi^3\over \pi^2(4f_\pi)^4 
\Lambda_\chi}\bigg\{{1\over u}
+2u-{8u^3\over 3}-{1+4u^2\over 4u^3}\ln(1+4u^2)\bigg\}.
\end{eqnarray}
Finally, the contribution, $V_{NN}^{\rm med,6}$, due to the three-body
contact interaction is momentum-independent and therefore does not
give rise to any noncentral components of the quasiparticle interaction.

\begin{figure*}
\begin{center}
\minipage{0.49\textwidth}
\includegraphics[scale=0.9,clip]{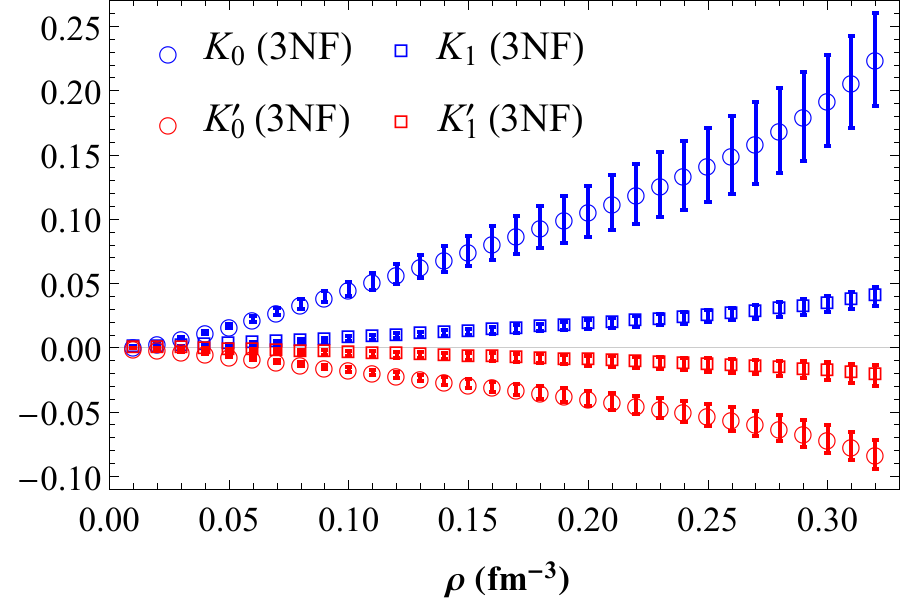}
\endminipage \hfill
\minipage{0.49\textwidth}
\includegraphics[scale=0.9,clip]{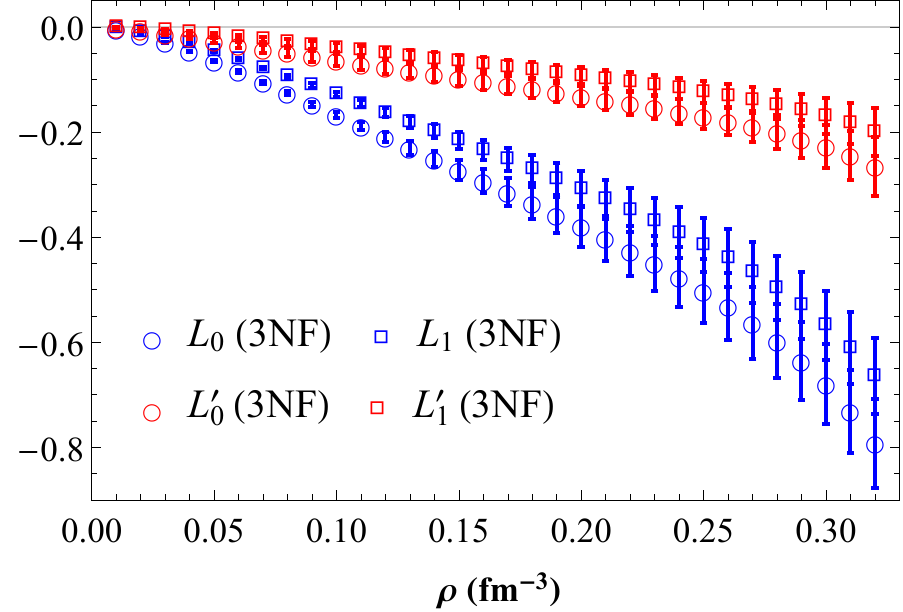}
\endminipage \hfill
\caption{First-order perturbative contributions to the dimensionless $L=0,1$ Fermi
liquid parameters of the center-of-mass tensor and cross-vector interactions as a 
function of density. Only three-body forces contribute at the Hartree-Fock level.}
\label{klfo}
\end{center}
\end{figure*}

In Figs.\ \ref{hfh} and \ref{klfo} we show as open symbols the $L=0,1$ Fermi liquid parameters
associated with the N2LO chiral three-body force as a function of density computed
at the Hartree-Fock level according to Eq.\ (\ref{qpi3nb}). All contributions vanish
in the $\rho \rightarrow 0$ limit and grow approximately linearly with the density up to
$\rho = \rho_0$. We find that in the case of the isotropic $H_0$ and $H_0^\prime$ Fermi 
liquid parameters, the three-body force enhances the contribution from two-body forces 
at all densities considered. In contrast, both $H_1$ and $H_1^\prime$ from two- and 
three-body forces exhibit large cancellations beyond nuclear saturation density. All three-body 
force contributions (except $V_{NN}^{\rm med,5}$, which is small) to the relative tensor 
Fermi liquid parameters obey the relationship $H_L = -3 H_L^\prime$ characteristic of 
one-pion exchange. At the Hartree-Fock level with two- and three-body forces, this 
relationship turns out to be an excellent approximation
relating the isoscalar and isovector relative tensor interactions. The center-of-mass tensor 
interaction from three-body forces is comparatively weak. The $L=0,1$ Fermi liquid parameters 
associated with both the isoscalar and isovector center-of-mass tensor force are all less than 
0.10 in magnitude at nuclear matter saturation density. We observe that the spin-nonconserving 
cross-vector interaction from the chiral three-body force is particularly strong in the isoscalar 
channel, with $L_0$ and $L_1$ from three-body forces reaching values around $-0.75$ at twice 
saturation density.


\subsection{Second-order perturbative contributions to Fermi liquid parameters}
\label{second}

We next consider the sum of the second-order particle-particle, particle-hole, and hole-hole
contributions to the noncentral Fermi liquid parameters. All are computed according to Eqs.\
(\ref{pp2})$-$(\ref{ph2}), except that a Hartree-Fock energy spectrum for the intermediate particle
and hole states is employed. This results in a reduction of the second-order contributions
by a density-dependent effective mass factor, which at saturation density
is on the order $M_{HF}^*/M \simeq 0.7$. Second-order perturbative contributions \cite{holt13prcb} to
the nucleon self-energy result in a slightly larger average effective mass on the order of
$M^*/M \simeq 0.85$ at nuclear saturation density. In the present study we neglect
such effects since the associated uncertainties are small compared to the choice of nuclear
force model.

\begin{figure}[t]
\begin{center}
\includegraphics[scale=0.9,clip]{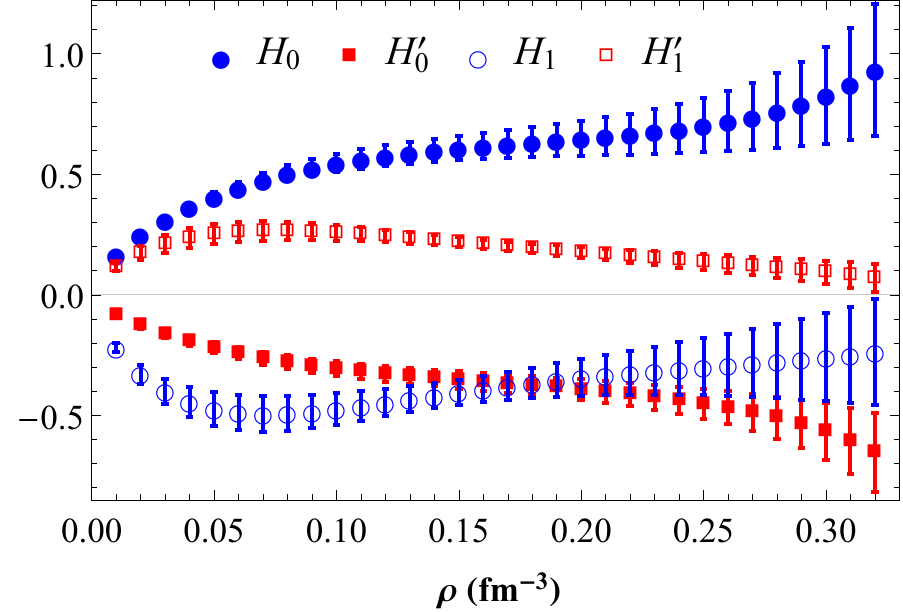}
\caption{Total $L=0,1$ Fermi liquid parameters of the relative tensor interaction
from two- and three-body forces as a function of the density. Error bars are obtained
from the standard deviation of the five chiral potentials considered in the present work.}
\label{hplot}
\end{center}
\end{figure}

\begin{figure}[t]
\begin{center}
\includegraphics[scale=0.9,clip]{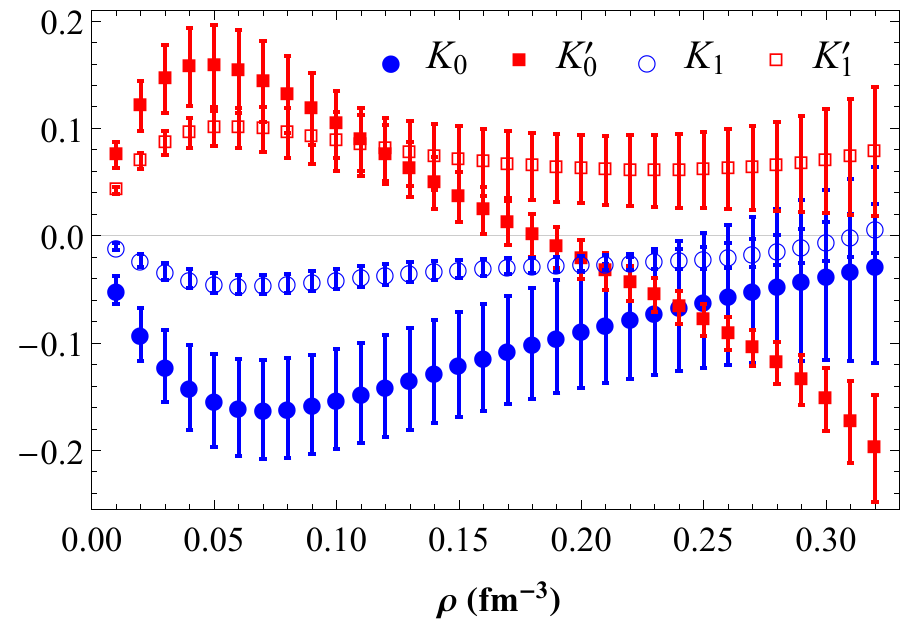}
\caption{Total $L=0,1$ Fermi liquid parameters of the center-of-mass tensor interaction
from two- and three-body forces as a function of the density. Error bars are obtained
from the standard deviation of the five chiral potentials considered in the present work.}
\label{kplot}
\end{center}
\end{figure}

\begin{figure}[t]
\begin{center}
\includegraphics[scale=0.9,clip]{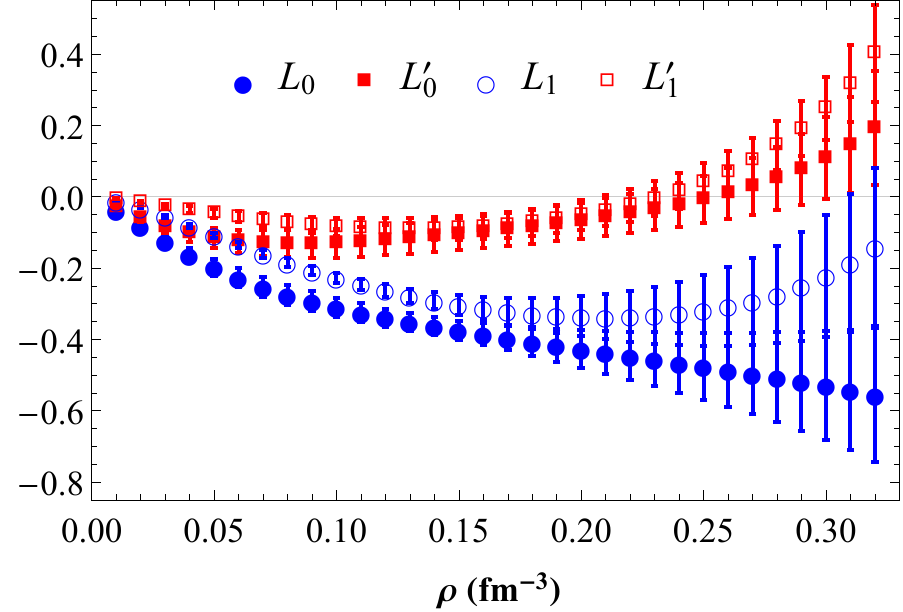}
\caption{Total $L=0,1$ Fermi liquid parameters of the cross-vector interaction
from two- and three-body forces as a function of the density. Error bars are obtained
from the standard deviation of the five chiral potentials considered in the present work.}
\label{lplot}
\end{center}
\end{figure}

In Figs.\ \ref{hplot}$-$\ref{lplot} we show the total dimensionless $L=0,1$ Fermi liquid 
parameters for the noncentral parts of the quasiparticle interaction as a function of 
density. This includes the first-order two-body and three-body contributions together with
the second-order particle-particle, hole-hole, and particle-hole diagrams.
Comparing the results to Figs.\ \ref{hfh} and \ref{klfo}, we see that overall
the second-order terms have a relatively small impact, in 
contrast to their large contributions to the central components \cite{holt12npa}. Neither
the particle-particle nor hole-hole diagram gives any contribution to the noncentral
Fermi liquid parameters larger than $0.1$ in magnitude across the range of densities
considered. The particle-hole diagram, 
however, gives contributions to the relative tensor and cross-vector Fermi liquid 
parameters on the order of $0.1-0.3$ in magnitude. None of the second-order diagrams 
lead to a sizable center-of-mass tensor interaction, and in Fig.\ \ref{kplot} we see
that this component of the quasiparticle interaction is very weak in both the isoscalar
and isovector channels in symmetric nuclear matter up to twice saturation density.

The $L=0,1$ Fermi liquid parameters of the isoscalar and isovector relative tensor force 
are very well constrained up to nuclear saturation density. In fact, the uncertainties on all four
parameters are less than $0.1$ in this regime but grow significantly beyond saturation
density. As in the case of one-pion exchange (OPE), the isotropic Landau parameters $H_0$ and
$H_0^\prime$ are respectively positive and negative across all densities considered. The 
generic relationship $H_L = -3H_L^\prime$, which is satisfied by OPE and most 3NF contributions, is 
violated due to second-order perturbative contributions. The corresponding tensor interactions
in the proton-neutron and proton-proton (neutron-neutron) channels are given by
\begin{eqnarray}
H^{pn}_L = H_L - H_L^\prime && \nonumber \\
H^{pp}_L = H^{nn}_L = H_L + H_L^\prime. &&
\end{eqnarray}
From Fig.\ \ref{hplot} we see that the combination of isoscalar and isovector tensor forces
produces a large proton-neutron effective interaction and a very small proton-proton 
(neutron-neutron) interaction, in agreement with a wide range of experimental data \cite{sagawa14}.

The Fermi liquid parameters of the cross-vector interaction are non-negligible in the
isoscalar channel. In particular, the value of $L_0$ (and to a lesser extent $L_1$) grows strongly 
with the density as a result of the N2LO chiral three-body force. The isovector cross-vector 
interaction, in contrast, remains small up to about twice saturation density. The large negative value 
of $L_0$ may be a concern in light of the normal stability conditions
\begin{equation}
{\cal C}_L > -(2L+1),
\label{stab}
\end{equation}
where ${\cal C} \in \{F,F^\prime,G,G^\prime\}$, for the central components of the quasiparticle
interaction. The presence of additional spin-dependent interactions $H, K, L$ 
(and $H^\prime, K^\prime, L^\prime$) that couple to $G$ (and $G^\prime$) modifies the stability 
criteria in Eq.\ (\ref{stab}), but to date only the effect of the relative tensor contributions have been 
considered \cite{backman79}.

\subsection{Central components of the quasiparticle interaction}
\label{cent}

In previous work \cite{holt12npa} we have computed the central Fermi liquid parameters in
symmetric nuclear matter including the effects of three-body forces. We update those results
to include theoretical uncertainties obtained by varying the chiral order and momentum-space
cutoff of the nuclear potential. In comparison to Ref.\ \cite{holt12npa} we also consider a 
larger range of densities in the present calculation. We then use standard relations 
\cite{migdal67,baym91} to study various nuclear observables that are directly related to the 
low-harmonic central Fermi liquid parameters. Since the tensor Fermi liquid parameters for 
symmetric nuclear matter are largely unconstrained by empirical data, benchmarking the 
central terms to empirical data is an important check on the nuclear force models and
many-body methods.

\begin{figure}[t]
\begin{center}
\includegraphics[scale=0.9,clip]{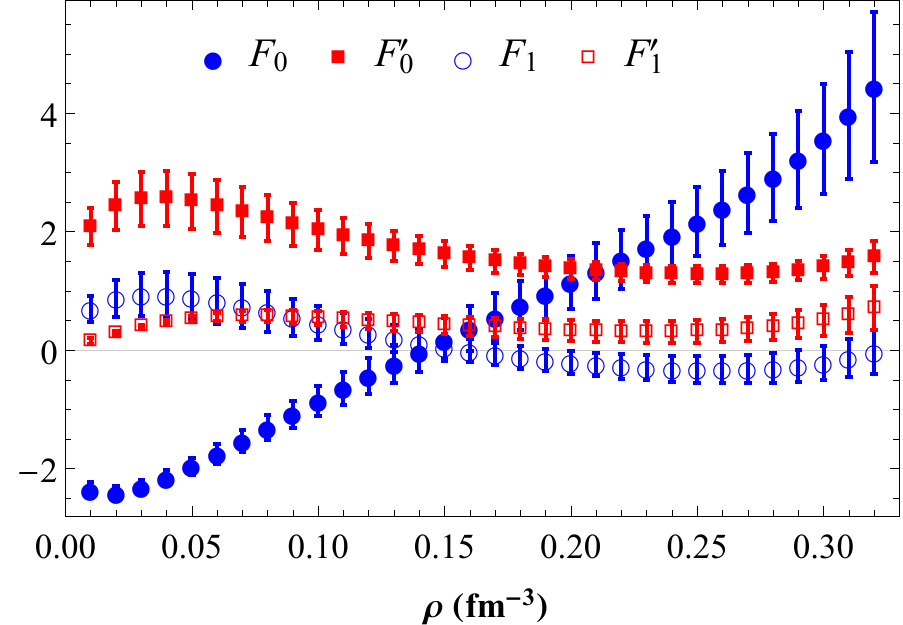}
\caption{Total $L=0,1$ Fermi liquid parameters of the spin-independent central parts of the
quasiparticle interaction
from two- and three-body forces as a function of the density. Error bars are obtained
from the standard deviation of the five chiral potentials considered in the present work.}
\label{fplot}
\end{center}
\end{figure}

\begin{figure}[t]
\begin{center}
\includegraphics[scale=0.9,clip]{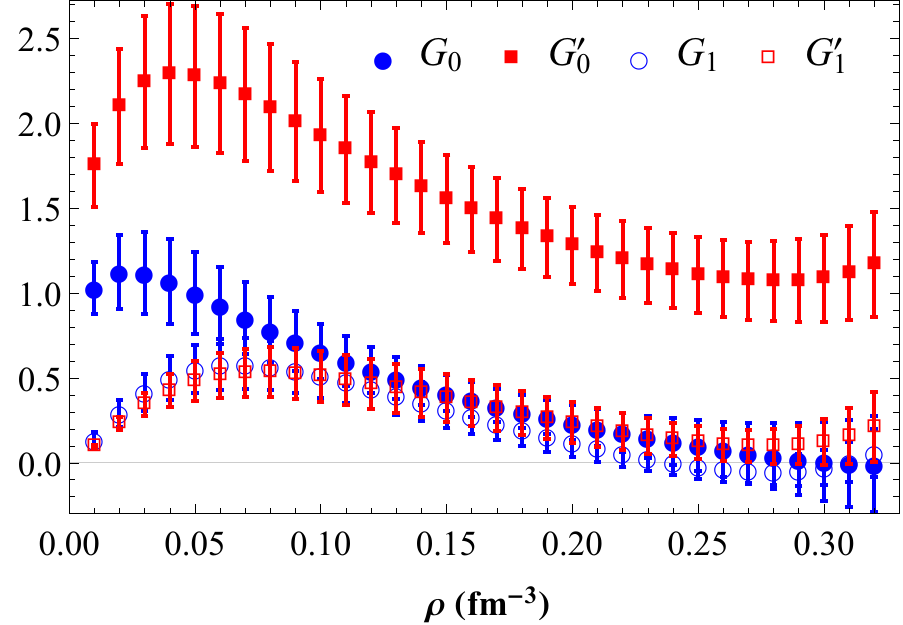}
\caption{Total $L=0,1$ Fermi liquid parameters of the spin-dependent central parts of the
quasiparticle interaction
from two- and three-body forces as a function of the density. Error bars are obtained
from the standard deviation of the five chiral potentials considered in the present work.}
\label{gplot}
\end{center}
\end{figure}

\begin{figure}[t]
\begin{center}
\includegraphics[scale=0.65,clip]{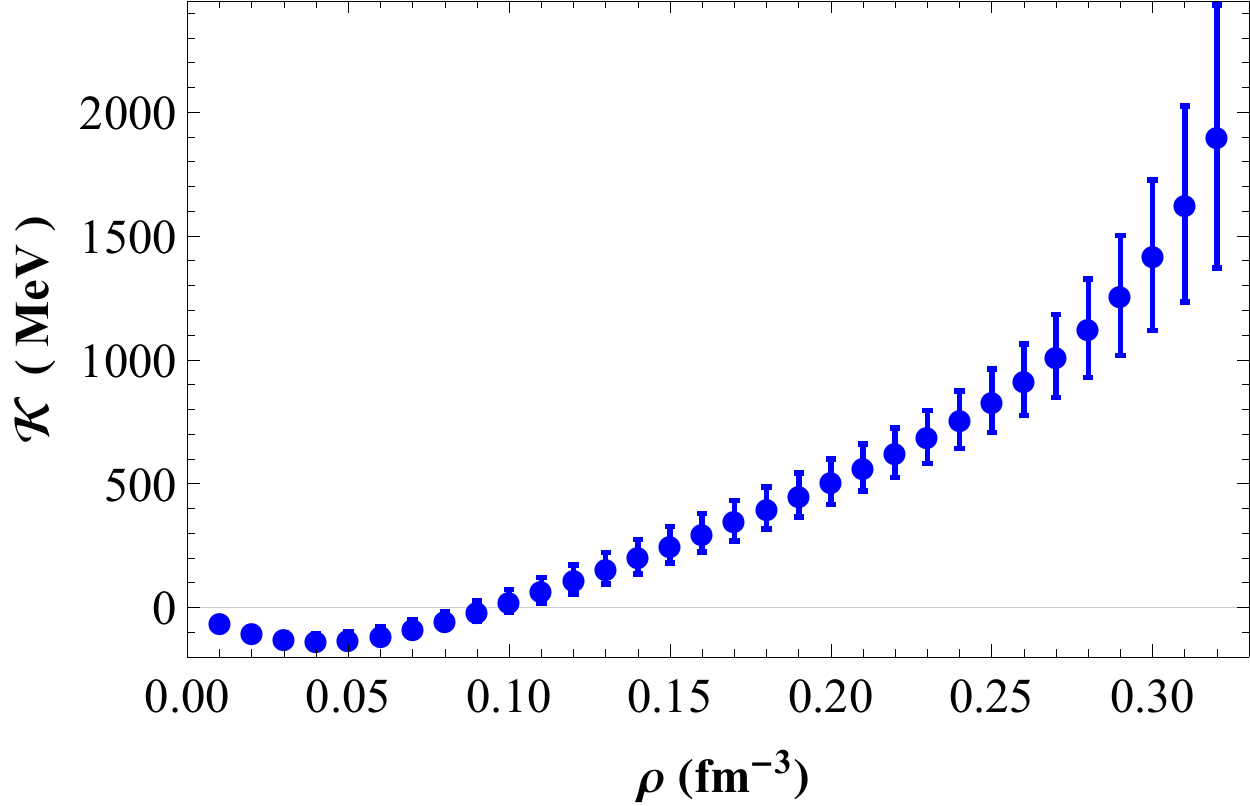}
\caption{Incompressibility of symmetric nuclear matter as a function of the density for
the two- and three-body chiral nuclear force models considered in the present work.}
\label{incomp}
\end{center}
\end{figure}

In Figs.\ \ref{fplot} and \ref{gplot} we show the $L=0,1$ Fermi liquid parameters associated
with the $F, F^\prime, G, G^\prime$ components of the quasiparticle interaction. The isotropic
spin- and isospin-independent Fermi liquid parameter $F_0$ is related to the nuclear 
incompressibility ${\cal K} = 9 \partial P / \partial \rho$,
where $P = \rho^2 \frac{\partial (E/A)}{\partial \rho}$, through
\be
{\cal K}=\frac{3k_F^2}{M^*} \left (1+F_0\right ).
\label{comp}
\ee
From Fig.\ \ref{fplot} we see that $F_0 < -1$ for $\rho \lesssim 0.10$\,fm$^{-3}$, which
corresponds to the well known instability of nuclear matter to density fluctuations associated with 
spinodal decomposition and cluster formation. However, the nearly linear dependence 
of $F_0$ on the nuclear density
results in a strongly increasing nuclear incompressibility, which we show in Fig.\ \ref{incomp}.
At $\rho=\rho_0$ the incompressibility lies in the range 
$190\,{\rm MeV} < {\cal K} < 380$\,MeV. While this is consistent with the empirical 
estimate of $220\,{\rm MeV} < {\cal K} < 260$\,MeV \cite{youngblood99,shlomo06}, the large
theoretical range is due to the fact the n2lo450, n2lo500, and n3lo500 nuclear forces 
saturate at too low of a density $\rho \simeq 0.14-0.15$\,fm$^{-3}$. 
In this case the contribution $\sim 18\rho \frac{\partial (E/A)}{\partial \rho}$ that is linear in the 
density strongly enhances the nuclear incompressibility.

\begin{figure}[t]
\begin{center}
\includegraphics[scale=0.65,clip]{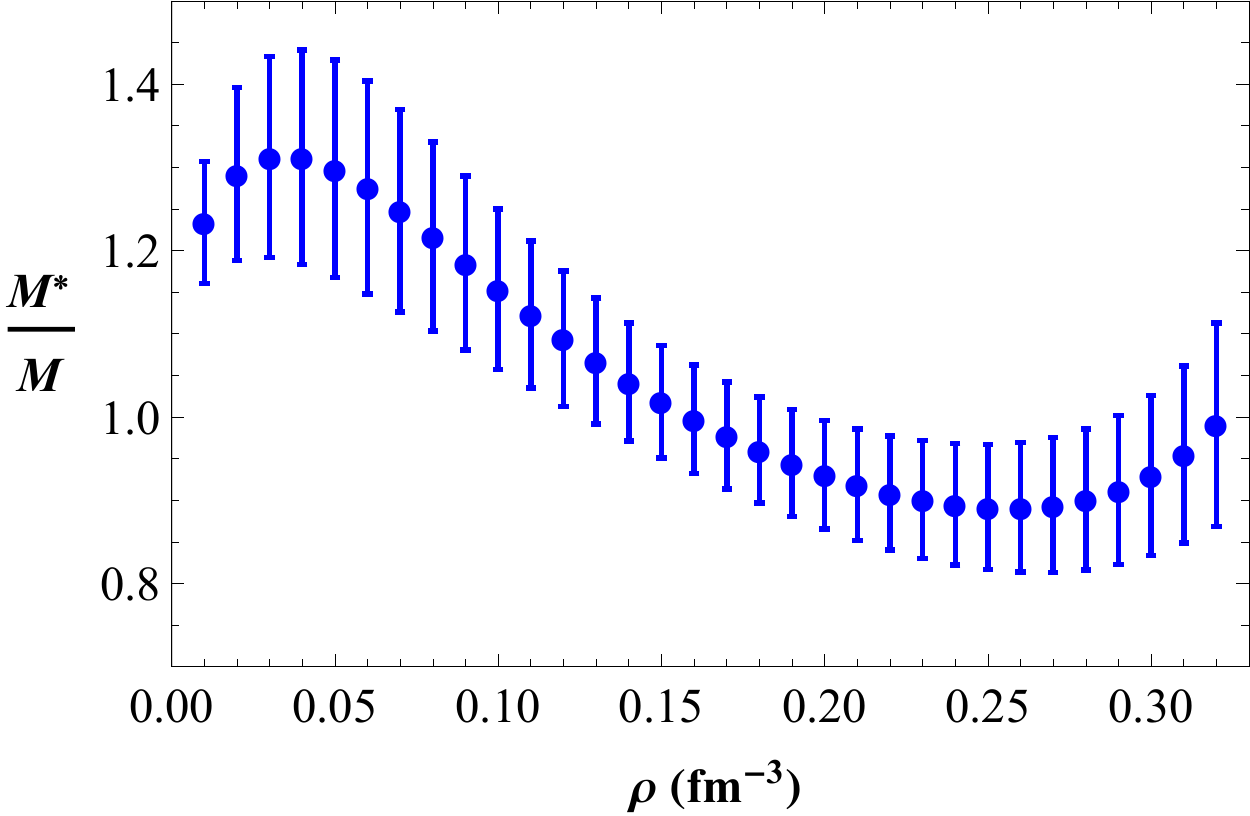}
\caption{Nucleon effective mass in symmetric nuclear matter as a function of the density for
the two- and three-body chiral nuclear force models considered in the present work.}
\label{mstar}
\end{center}
\end{figure}

In Eq.\ (\ref{comp}) the quasiparticle effective mass $M^*$ is related to the Landau parameter
$F_1$ through
\be
\frac{M^*}{M} = 1+\frac{F_1}{3},
\label{effmass}
\ee
with $M=938.9182$\,MeV the average nucleon mass. From Fig.\ \ref{fplot} we find
that $F_1 > 0$ for $\rho \lesssim \rho_0$ and consequently the effective mass 
is larger than the free-space mass. For $\rho \gtrsim \rho_0$ the effective mass is
typically less than that of a free nucleon, but the decrease in the effective mass with
increasing density is not nearly as large as in most mean field models and other 
approaches to scaling masses \cite{friman96} at first order.
The large effective mass is due almost solely to the second-order 
particle-hole diagram, which gives a contribution $F^{(2ph)}_1 \simeq 1$ for all densities
up to $\rho = 2\rho_0$. From the study of nuclear level densities in the vicinity of the Fermi surface, 
the effective mass has been estimated \cite{brown63,bertsch68} to lie 
close to that of a free nucleon $M^* \simeq M$. In Fig.\ \ref{mstar} we show the effective mass 
as a function of density together with the theoretical uncertainty estimates. At saturation density 
we find the range $0.9 < M^*/M < 1.1$. Since $M^*/M \rightarrow 1$ as
$\rho \rightarrow 0$, the effective mass must rise rather quickly at low densities.

\begin{figure}[t]
\begin{center}
\includegraphics[scale=0.65,clip]{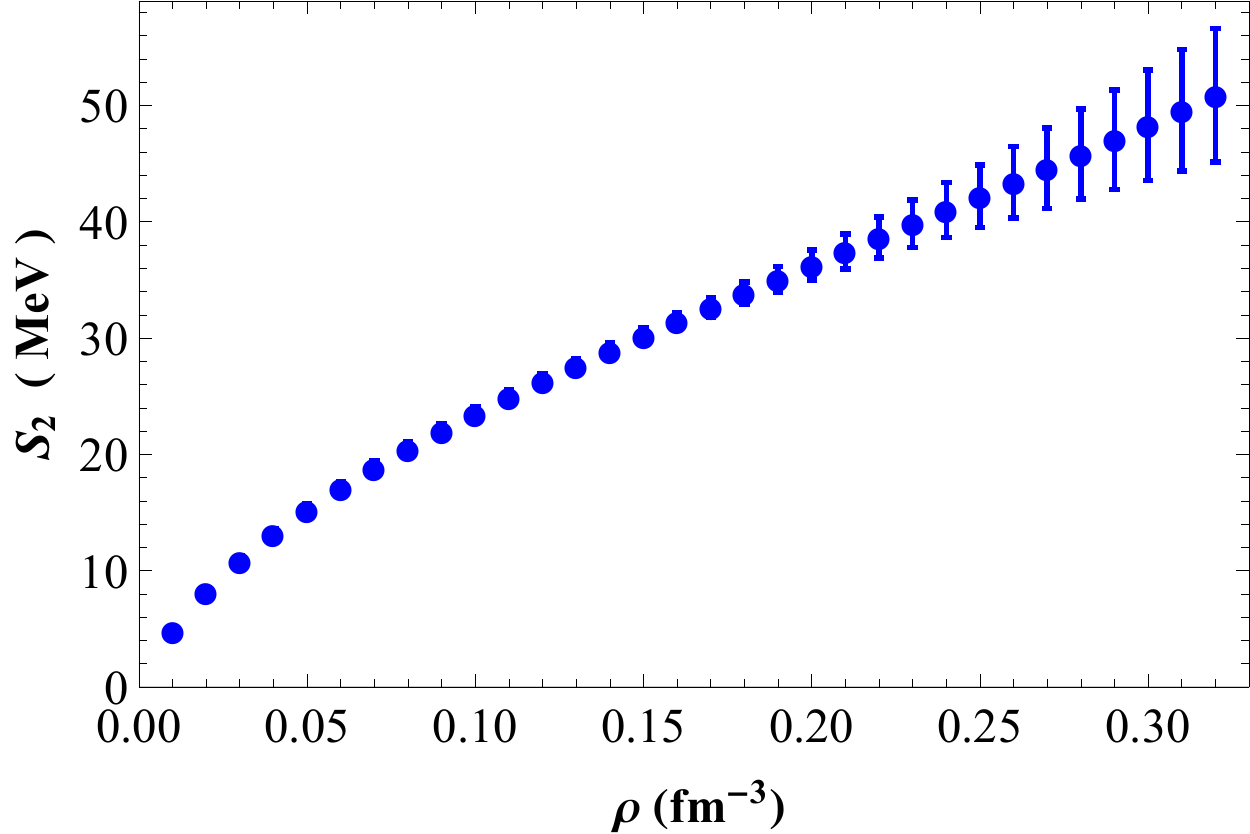}
\caption{Isospin-asymmetry energy as a function of the density for
the two- and three-body chiral nuclear force models considered in the present work.}
\label{syme}
\end{center}
\end{figure}

We define the density-dependent isospin-asymmetry energy $S_2(\rho)$ 
as the coefficient of the quadratic term in an expansion
of the energy per particle of isospin-asymmetric nuclear matter in powers of the parameter 
$\delta_{np} = \frac{\rho_n-\rho_p}{\rho_n+\rho_p}$:
\be
\frac{E}{A}(\rho,\delta_{np}) = \frac{E}{A}(\rho,0) + S_2(\rho)\delta_{np}^2 + \cdots
\label{esyme}
\ee
Generically \cite{kaiser15,wellenhofer16} the energy per particle contains non-analytic 
contributions in $\delta_{np}$ beyond the quadratic term in Eq.\ (\ref{esyme})
 when second-order perturbative corrections are included in the equation of state, 
 but at low temperatures it is nevertheless a good 
approximation to retain only the quadratic term in the expansion in Eq.\ (\ref{esyme}). 
The isospin-asymmetry energy is related to the isotropic part of the $F^\prime$ 
contribution to the quasiparticle interaction:
\be
S_2=\frac{k_F^2}{6M^*} \left (1+F_0^\prime \right ).
\label{comp}
\ee
In Fig.\ \ref{syme} we plot $S_2(\rho)$ and associated uncertainties up to twice nuclear
matter saturation density. We find the peculiar feature that the variations in the 
Landau parameter $F_1$ (which enters into the definition of the effective mass $M^*$) and 
in the Landau parameter $F_0^\prime$ are correlated in such a way as to produce a very small
error band for the isospin-asymmetry energy up to nuclear saturation density. For instance, 
at nuclear matter saturation density, we obtain $30\,{\rm MeV} < S_2(\rho_0) < 32\,{\rm MeV}$,
which is consistent with other recent microscopic uncertainty estimates \cite{holt17prc,drischler17}
but with a much smaller error band.
It is not clear what could lead to the correlation between $F_1$ and $F_0^\prime$, and therefore
we tentatively attribute the very small errors in $S_2(\rho)$ to a chance cancellation.

From Fig.\ \ref{gplot} we see that $G_0^\prime$ remains large and positive for all densities 
considered. At nuclear saturation density, we find $1.2 < G_0^\prime < 1.8$, which is
consistent with extractions \cite{bertsch81,gaarde81} from fitting the peak energy of giant Gamow-Teller
resonances in heavy nuclei. Such fits give a range $1.4 < G_0^\prime < 1.6$ \cite{bender02} 
but rely on certain model assumptions related to the shape of the parametrized single-particle
potential and the form of the effective interaction. In addition the authors of Ref.\ \cite{bender02} 
find correlations between the value of $G_1^\prime$ and the position of the 
energy peak of the Gamow-Teller resonance when $G_0^\prime$ is kept fixed, leading to
further uncertainties in the extraction of $G_0^\prime$ from resonance data.

\begin{figure}[t]
\begin{center}
\includegraphics[scale=0.35,clip]{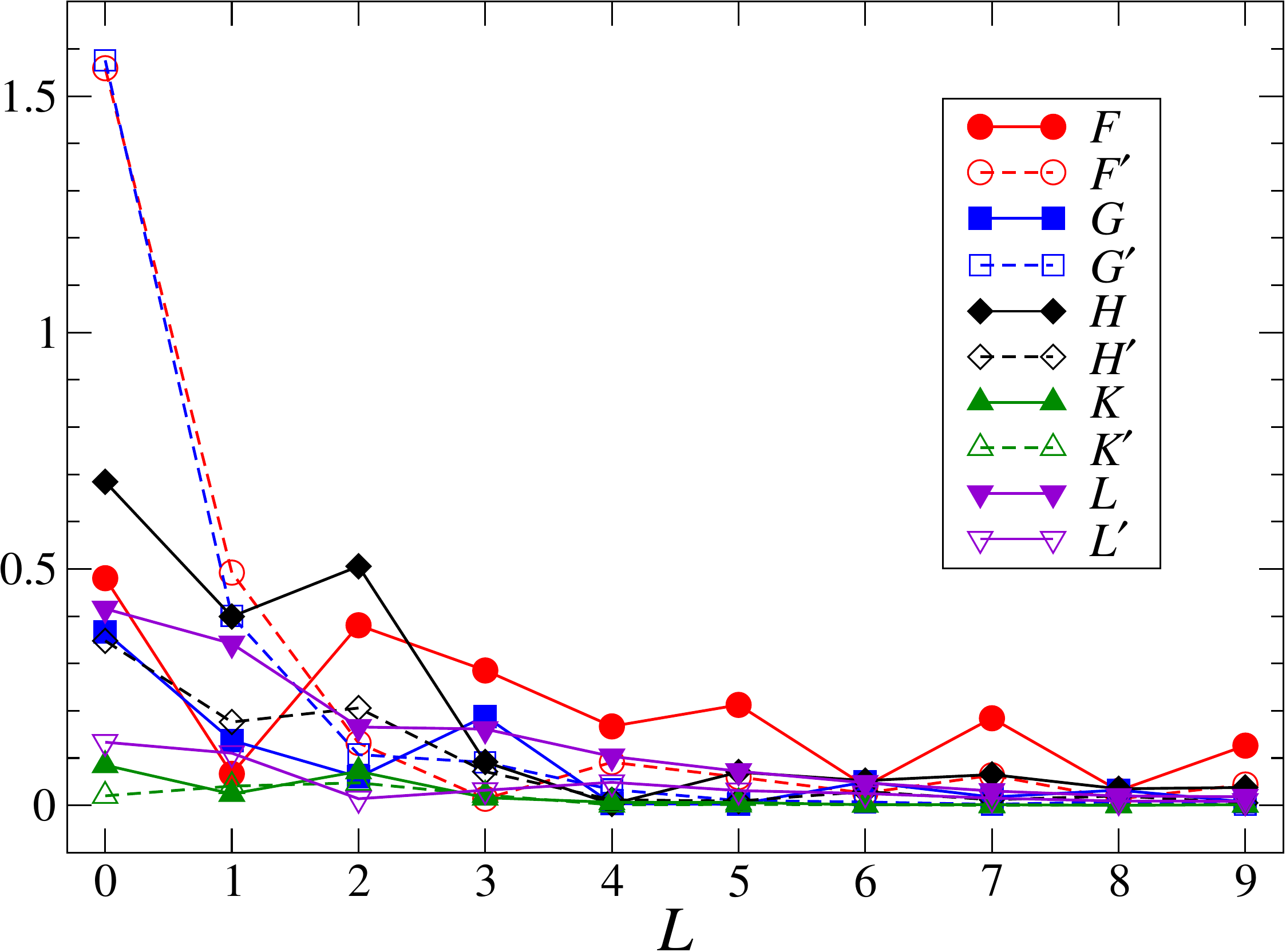}
\caption{Convergence of the Legendre polynomial expansion for each of the 
different contributions to the quasiparticle interaction. Results are shown only for
n3lo450 two- and three-body forces at nuclear matter saturation density.
Terms up to $L=9$ are considered.}
\label{conv}
\end{center}
\end{figure}

Finally, we investigate the convergence of the Legendre polynomial decomposition, 
Eq.\ (\ref{gflp}), for the noncentral components of the quasiparticle interaction. From
Figs.\ \ref{hplot}--\ref{lplot} we see that in some cases the $L=1$ Fermi liquid parameters
are comparable in magnitude to the $L=0$ parameters at nuclear matter saturation
density. Using the tensor parametrizations in Eq.\ (\ref{qpi}), it is expected \cite{olsson04} 
that the convergence is much improved compared to alternative choices, such as
\begin{equation}
H(\vec p_1,\vec p_2) S_{12}(\hat q) = \frac{q^2}{k_f^2} \tilde H(\vec p_1,\vec p_2) S_{12}(\hat q)
\end{equation}
employed in Refs.\ \cite{haensel75,brown77,haensel82}. In Fig.\ \ref{conv} we plot the
ten lowest dimensionless Fermi liquid parameters from the n3lo450 potential at 
nuclear matter saturation density. It is clear that the slowest convergence is in the spin- and 
isospin-independent part of the quasiparticle interaction $F$, which even up to $L=9$ 
has contributions greater than 0.1. The Legendre polynomial expansion in all other 
channels is nearly converged by $L=5$.


\section{Conclusions and outlook}

In the present work we have computed for the first time the full set of central and noncentral
contributions to the quasiparticle interaction in symmetric nuclear matter up to twice
nuclear saturation density. We have derived general formulas that allow one to extract the 
associated scalar functions from appropriate linear combinations of spin- and isospin-space 
matrix elements. Both two- and three-body forces are included at first- and second-order in 
perturbation theory, with the involved numerical calculations of the second-order diagrams
benchmarked against model interactions. 
Three-body forces at the Hartree-Fock level are shown to give important 
contributions to the relative tensor and cross-vector interactions. Indeed, the isovector
cross-vector interaction is dominated by three-body forces, in particular the two-pion exchange
term proportional to the low-energy constant $c_4$, and only the second-order 
particle-hole diagram leads to a modest reduction of the strength in this channel. While the 
relative tensor force from the free-space nucleon-nucleon interaction is enhanced in
the medium by three-body forces and second-order perturbative corrections, the 
center-of-mass tensor force remains relatively weak.

We have considered five different nuclear force models in order to estimate theoretical
uncertainties. Up to nuclear saturation density, the relative tensor force in both the isoscalar
and isovector channels is well constrained
by microscopic many-body theory, which should be valuable for efforts to include effective
tensor forces in mean-field modeling and density functional theory. In addition we find robust 
evidence for a strong isovector cross-vector interaction which is not normally included in 
mean-field models and may be important for spin-dependent phenomena. We benchmark 
the quality of our results against bulk nuclear matter properties, such as the incompressibility,
isospin-asymmetry energy, and nucleon effective mass, which are directly related to selected 
central Fermi liquid parameters. While the theory uncertainties are sometimes large, 
in all cases the results are consistent with empirical constraints.

The present calculations are a step toward the microscopic description of response functions 
in nuclear matter consistent with equations of state exhibiting realistic nuclear saturation properties. 
In the future this work will be extended to isospin-asymmetric nuclear systems with applications 
to neutrino processes in neutron stars and supernovae.

\acknowledgments
T.\ R.\ Whitehead thanks S.\ Guha for informative discussions. Work supported by the 
National Science Foundation under Grant No.\ PHY1652199. The work of N.\ Kaiser 
has been supported in part by DFG and NSFC through funds provided to the Sino-German 
CRC 110 ``Symmetries and the Emergence of Structure in QCD". Portions of this research 
were conducted with the advanced computing resources provided by Texas A\&M 
High Performance Research Computing. 


\vspace{.3in}

\section{Appendix}

\subsection{Central components of the quasiparticle interaction
from the N2LO chiral three-body force}

In this section we provide for completeness the central Fermi liquid parameters for 
arbitrary values of $L$ resulting from the N2LO chiral three-body force. The 
various terms are organized according to the different topologies shown in 
Fig.\ \ref{mednn}. Some contributions are explicitly separated into a direct ``d'' and 
exchange ``e'' term. For notational convenience we 
introduce  the abbreviations $\sigma = \vec \sigma_1\cdot \vec \sigma_2$ 
and $\tau = \vec \tau_1\cdot \vec \tau_2$. The contributions from the six topologies
read

\begin{widetext}

\begin{equation}
{\cal F}_L^{(\rm med,1)} =({\bf \sigma}-3)(3-{\bf 
\tau}){4g_A^2m_\pi^3u \over
9\pi^2f_\pi^4} \int_0^u\!dx\,x^3 (2L+1)P_L(1-2x^2u^{-2})\,{c_1+2c_3 
x^2\over
(1+4x^2)^2}\,,
\end{equation}

\begin{eqnarray}
{\cal F}_L^{(\rm med,2)} &=&(3-{\bf \sigma})(3-{\bf 
\tau}){g_A^2m_\pi^3 \over
(12\pi)^2f_\pi^4u^5}\int_0^u\!dx\,{x^3(2L+1)\over 1+4x^2} P_L(1-2x^2u^{-2})
\bigg\{16u^3(c_3-2c_4)\arctan 2u\nonumber \\ && +\bigg[{3x^2\over 
2}(c_3+c_4)
(4+u^{-2})+6u^2(c_4-c_3-2c_1)-3c_1-{c_3+c_4\over 2}\bigg] 
\ln(1+4u^2)\nonumber
\\ &&  +2(c_3+c_4)x^2(8u^4-6u^2-3)+12c_1 u^2(1+2u^2)\nonumber \\ &&
+2c_3 u^2\bigg(1-6u^2+{8u^4\over 3}\bigg)+2c_4 u^2\bigg(1+18u^2
-{40u^4\over 3}\bigg)\bigg\} \,,
\end{eqnarray}

\begin{equation}
{\cal F}_0^{(\rm med,3d)} = {g_A^2m_\pi^3\over 
16\pi^2f_\pi^4}\bigg\{24u(c_3-c_1)
-8c_3 u^3+6(6c_1-5c_3)\arctan 2u +{3\over u}(3c_3-4c_1)\ln(1+4u^2) \bigg\} \,,
\end{equation}

\begin{eqnarray}
{\cal F}_L^{(\rm med,3e)} &=&{3 g_A^2m_\pi^3 \over 
32\pi^3f_\pi^4}\int_0^u\!dl\,
l^2 \int_{-1}^1\!dx \int_{-1}^1\!dy \int_0^\pi\!d\phi\, 
{(2L+1)P_L(z)\over (1+u^2
+l^2-2u l x)(1+u^2+l^2-2u l y)}\nonumber \\ && \times \Big\{(1+{\bf \sigma})
(1+{\bf \tau})\Big[2c_1 (u^2z-u l x-u l y+l^2) +c_3 (u^2z-u l x-u l 
y+l^2)^2\Big]
\nonumber \\ && + (3-{\bf \sigma})(3-{\bf \tau}){c_4\over 9}\Big[(u^2z-u l x
-u l y+l^2)^2- (u^2+l^2-2u l x)(u^2+l^2-2u l y)\Big] \bigg\}  \,,
\end{eqnarray}
with $z = xy +\sqrt{(1-x^2)(1-y^2)}\cos\phi$,

\begin{equation}
{\cal F}_L^{(\rm med,4)} =(3-{\bf \sigma})(3-{\bf \tau}){g_A c_D m_\pi^3u
\over 18 \pi^2f_\pi^4\Lambda_\chi } \int_0^u\!dx\,{x^3 (2L+1) \over 1+4x^2}
P_L(1-2x^2u^{-2})\,,
\end{equation}
to which only the exchange term contributes,

\begin{equation}
{\cal F}_0^{(\rm med,5)} =(3-{\bf \sigma}-{\bf \tau}- {\bf \sigma \tau})
{g_A c_D m_\pi^3 \over 16 \pi^2f_\pi^4\Lambda_\chi }\bigg\{{2u^3 \over 3}-u
+\arctan 2u-{1\over 4u} \ln(1+4u^2)\bigg\}\,,
\end{equation}

\begin{equation}
{\cal F}_0^{(\rm med,6)} =({\bf \sigma}+{\bf \tau}+{\bf \sigma \tau}-3)
{c_E\,k_f^3 \over 4\pi^2f_\pi^4\Lambda_\chi }\,.
\end{equation}

\end{widetext}


\clearpage 
 
\bibliographystyle{apsrev4-1}

\end{document}